\documentclass[journal=jacsat,manuscript=article]{achemso}

\usepackage{chemformula}
\usepackage[T1]{fontenc}
\usepackage{amsmath,amssymb}
\usepackage{graphicx}
\usepackage{xcolor}
\usepackage{wrapfig}
\usepackage{booktabs}
\usepackage{siunitx}
\usepackage{subcaption}

\usepackage[caption=false]{subfig}

\newcommand{\activepoly}{pTPAC$_6$DTP}
\newcommand{\polymer}{poly(4-((6-(4H-dithieno[3,2-b:2',3'-d]pyrrol-4-yl)hexyl)oxy)-N,N-diphenylaniline)}

\newcommand{\degC}{$^{\circ}$C}
\newcommand{\comm}[1]{}
\definecolor{grey}{rgb}{0.9, 0.9, 0.9}

\author{Stephen H. Foulger}
\affiliation[COMSET]{Center for Optical Materials Science and Engineering Technologies (COMSET), Clemson University, Clemson, SC 29634 USA}
\alsoaffiliation[MSE]{Department of Materials Science and Engineering, Clemson University, Clemson, SC 29634 USA}
\alsoaffiliation[BioE]{Department of Bioengineering, Clemson University, Clemson, SC 29634 USA}
\email{foulger@clemson.edu}
\phone{(864) 656-1045}

\author{Yuriy Bandera}
\affiliation[COMSET]{Center for Optical Materials Science and Engineering Technologies (COMSET), Clemson University, Clemson, SC 29634 USA}
\alsoaffiliation[MSE]{Department of Materials Science and Engineering, Clemson University, Clemson, SC 29634 USA}

\author{Igor Luzinov}
\affiliation[COMSET]{Center for Optical Materials Science and Engineering Technologies (COMSET), Clemson University, Clemson, SC 29634 USA}
\alsoaffiliation[MSE]{Department of Materials Science and Engineering, Clemson University, Clemson, SC 29634 USA}

\author{Travis Wanless}
\affiliation[COMSET]{Center for Optical Materials Science and Engineering Technologies (COMSET), Clemson University, Clemson, SC 29634 USA}
\alsoaffiliation[MSE]{Department of Materials Science and Engineering, Clemson University, Clemson, SC 29634 USA}

\author{Lubomir Kostal}
\affiliation[CAS01]{Institute of Physiology, Computational Neuroscience Group, Czech Academy of Sciences, 142 20 Prague 4, Czech Republic}

\author{Vojtech N\'{a}da\v{z}dy}
\affiliation[SCS]{Institute of Physics Slovak Academy of Sciences, 845~11 Bratislava, Slovak Republic}

\author{Petr Janovsk\'{y}}
\affiliation[TBU1]{Department of Chemistry, Faculty of Technology, Tomas Bata University in Zl\'{i}n, 760 01 Zl\'{i}n, Czech Republic}

\author{Jarmila Vil\v{c}\'{a}kov\'{a}}
\affiliation[TBU2]{Department of Physics and Materials Engineering, Faculty of Technology,
Tomas Bata University in Zl\'{\i}n, 760 01 Zl\'{\i}n, Czech Republic}
\alsoaffiliation[CPS]{Centre of Polymer Systems, Tomas Bata University in Zl\'{\i}n, 760 01 Zl\'{\i}n, Czech Republic}

\title[title]{Polymer-based probabilistic bits for thermodynamic computing}

\abbreviations{DOS, PDOS, OPDOS}

\begin{document}

\maketitle
\clearpage   

\begin{abstract}
Probabilistic bits (p-bits) are stochastic hardware elements whose output probability can be tuned by an input bias, offering a route to energy-efficient architectures that exploit, rather than suppress, fluctuations. Here we report p-bit generation in an organic memristive device, establishing polymers as the first class of soft-matter systems to realize probabilistic hardware. The active element is a dithieno[3,2\textendash b:2',3'\textendash d]pyrrole (DTP)–backbone polymer with pendant triphenylamine (TPA) groups, whose stochastic resistance fluctuations are converted into binary outputs by a simple voltage-divider / comparator circuit. The resulting probability distributions follow logistic transfer functions, characteristic of stochastic binary neurons. Separately, ensembles of pulsed $I$–$V$ measurements were analyzed to construct binned current distributions, from which the discrete Shannon entropy was calculated. Peaks in this entropy coincide with bias conditions that maximize variability in the memristor voltage drop, directly linking device-level stochasticity to intrinsic material properties. Dielectric analysis shows that pendant TPA units provide dynamically active relaxation modes, while energy-resolved electrochemical impedance spectroscopy and density functional theory calculations indicate that the frontier orbitals of DTP, TPA and ITO align within the transport gap to produce a bifurcated percolation network. The correspondence between microscopic relaxation dynamics, electronic energetics and macroscopic probabilistic response highlights how organic semiconductors can serve as chemically tunable entropy sources, opening a polymer-based pathway toward thermodynamic computing.

\end{abstract}

\section{Introduction}
Digital logic architectures remain fundamentally inefficient for probabilistic inference, combinatorial optimization, and other tasks that benefit from stochastic search strategies\cite{Gaines1969,Alaghi2013}. To address this gap, the computing industry is moving toward thermodynamic and probabilistic hardware, with application-specific integrated circuits (ASICs) designed to harness physical noise and fluctuations for energy-efficient AI acceleration\cite{Freitas2021,Moore2025,normalcomputing2025}. 
This shift reflects a broader trend toward physics-based circuit design, where randomness is exploited rather than suppressed as a computational resource\cite{Camsari2017}. 
Within this framework, probabilistic bits (p-bits), binary elements whose output probability is tunable by a circuit parameter, have emerged as promising building blocks\cite{Camsari2017,borders2019}. 
Freitas and co-workers further strengthened this perspective by showing, within a stochastic thermodynamics framework, that p-bits can be modeled as thermodynamically consistent devices that harness thermal noise for computation\cite{Freitas2021}. While magnetic tunnel junctions\cite{borders2019,Sutton2017} and CMOS-based stochastic circuits\cite{palem:2005} provide established foundations, the p-bit concept has more recently expanded into optical implementations\cite{Carmes2023}. In parallel, organic and polymer platforms offer unique advantages, notably structural tunability and compatibility with versatile, low-cost fabrication methods such as printing\cite{Baeg2013,Sonder2012}, making them attractive candidates for scalable p-bit hardware.

\begin{wrapfigure}{r}{50mm}
\centering
    \includegraphics[trim={1mm 1mm 1mm 1mm}, clip, width=40mm]{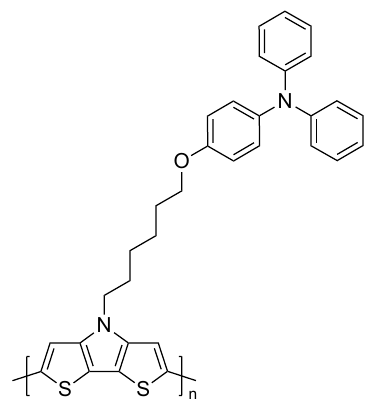}
\caption{\footnotesize{Structure of the repeat unit for \polymer\ (\activepoly), neutral form. The oxidized form was employed in all device testing.}}
\label{fig:yb1518}
\end{wrapfigure}

Here, we present a materials-focused approach based on polymeric memristors that leverage intrinsic stochasticity arising from coupled redox processes and conformational dynamics in soft matter systems\cite{Matsui2018,grant2022,grant2023_01,grant2023_02}. 
Using a triphenylamine-substituted dithieno[3,2-b:2',3'-d]pyrrole (DTP) polymer (\polymer, \activepoly), we demonstrate voltage-gated p-bit behavior with tunable entropy output (cf.\ \textbf{Figure~\ref{fig:yb1518}}). DTP-derived polymers have already been deployed in diverse technologies, including energy storage\cite{jm:dtp01}, photovoltaics\cite{ez:dtp01}, OLEDs\cite{se:dtp01}, thin-film transistors\cite{gl:dtp01}, biosensors\cite{ha:dtpredox01}, electrochromic systems\cite{ko:dtp01}, and bistable memory\cite{cw:dtp01}. 
The fused, electron-rich dithieno[3,2-b:2',3'-d]pyrrole (DTP) core promotes backbone planarity and enhances $\pi$–$\pi$ overlap, thereby extending effective conjugation along the poly(DTP) backbone.
Our earlier report\cite{foulger:2025_01} provided the first demonstration of a polymer memristor operating as a probabilistic bit, and the present study advances this foundation by delivering the first detailed mechanistic analysis of dynamic response, directly linking intrinsic material properties to stochastic bit generation and establishing a framework for polymer-based probabilistic hardware.

We approach the problem through three interconnected perspectives. At the circuit level, we demonstrate that a single \activepoly\ device spans the stochastic–deterministic continuum via a bias-controlled logistic transfer function. At the device level, we confirm memristive identity through frequency-dependent pinched hysteresis and analyze the molecular motions that set the relevant timescales. At the molecular level, 
electrochemical reduction--electrochemical impedance spectroscopy (ER--EIS)
and quantum calculations reveal a backbone/pendant bifurcation of transport pathways, linking electronic structure to probabilistic function.

\section{Results}

\begin{wrapfigure}{r}{50mm}
\centering
    \includegraphics[trim={1mm 1mm 1mm 1mm}, clip, width=40mm]{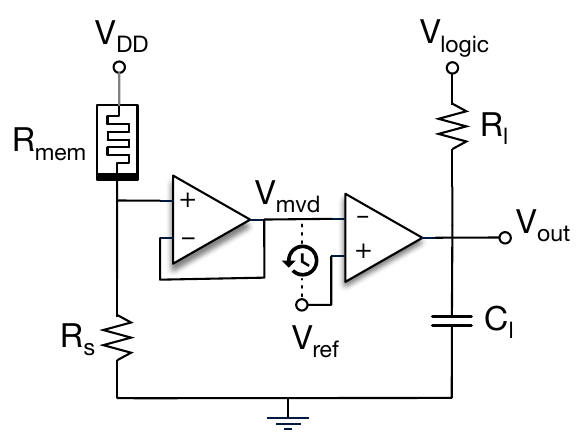}
\caption{\footnotesize{Circuit schematic of the p-bit primitive. A memristive element $R_{\text{mem}}$ in series with $R_{s}$ forms a voltage divider that generates $V_{\text{mvd}}$, which is compared against a reference voltage $V_{\text{ref}}$ to produce the stochastic output. The probability of generating $V_{\text{out}}$ follows the sigmoidal transfer function in Eq.~\ref{eq:stochastic_binary_neuron}, with $R_{l}$ and $C_{l}$ setting the load and output dynamics.}}
\label{fig:circuit}
\end{wrapfigure}
The sigmoidal transfer function is a cornerstone of artificial intelligence (AI) and machine learning (ML) architectures\cite{hornik1989,rumelhart1986,Bengio2013}.  
In artificial neural networks, it provides a smooth, nonlinear mapping between input and output, enabling the system to capture complex dependencies that cannot be represented by linear functions alone.  
Its bounded output in the range $[0,1]$ makes it directly interpretable as a probability, which is particularly valuable for classification tasks and probabilistic inference.  
Moreover, the differentiability of the sigmoidal function ensures compatibility with gradient-based optimization methods, such as backpropagation, which are central to modern ML training algorithms.  
In binary stochastic neurons, the sigmoidal response governs the probability of a unit adopting state ``0'' or ``1,'' embedding stochasticity directly into the computational process.  

The p-bit primitive employed in this work leverages this principle at the hardware level.  
Unlike software-based neurons, which implement stochasticity through pseudo-random number generators, 
the p-bit achieves an inherent  probabilistic response 
through its physical circuit properties.  
The circuit presented in  \textbf{Figure~\ref{fig:circuit}} follows a transfer function analogous to that of an ideal binary stochastic neuron, thereby enabling hardware-native sampling and probabilistic inference that are essential for energy-efficient probabilistic computing.  
The present circuit design builds on earlier implementations~\cite{foulger:2025_01} but eliminates the need for an n-type metal–oxide–semiconductor (NMOS) transistor to modulate $V_{\text{mvd}}$. These circuits trace back to architectures originally developed for generating p-bits with stochastic magnetic tunnel junctions~\cite{borders2019}. The modification retains the same underlying principles of harnessing stochastic behavior for p-bit generation while integrating them into a new material platform.
In this circuit, the reference voltage (\(V_{\text{ref}}\)) is generated by adding a small offset voltage (\(V_{\text{offset}}\)) to a previously measured \(V_{\text{mvd}}\), which is periodically updated every 10 cycles to correct for any drift in the memristor's behavior over time.
The comparator then compares the current \(V_{\text{mvd}}\) with this reference voltage (\(V_{\text{ref}} = V_{\text{mvd}} + V_{\text{offset}}\)). 
The comparator's output (\(V_{\text{out},i}\)) is either 0 V (logic low) or +5 V (logic high), indicating whether the current \(V_{\text{mvd}}\) is 
above or below the offset-adjusted \(V_{\text{ref}}\). 
By adjusting the offset voltage, the probability of generating a ``0'' (\(P_0\))
or a ``1'' (\(P_1\)) is controlled. 

\subsection*{Stochastic Transfer Function and P-Bits}

\begin{figure}[t]
\centering
    \includegraphics[trim={3mm 3mm 3mm 3mm}, clip, width=90mm]{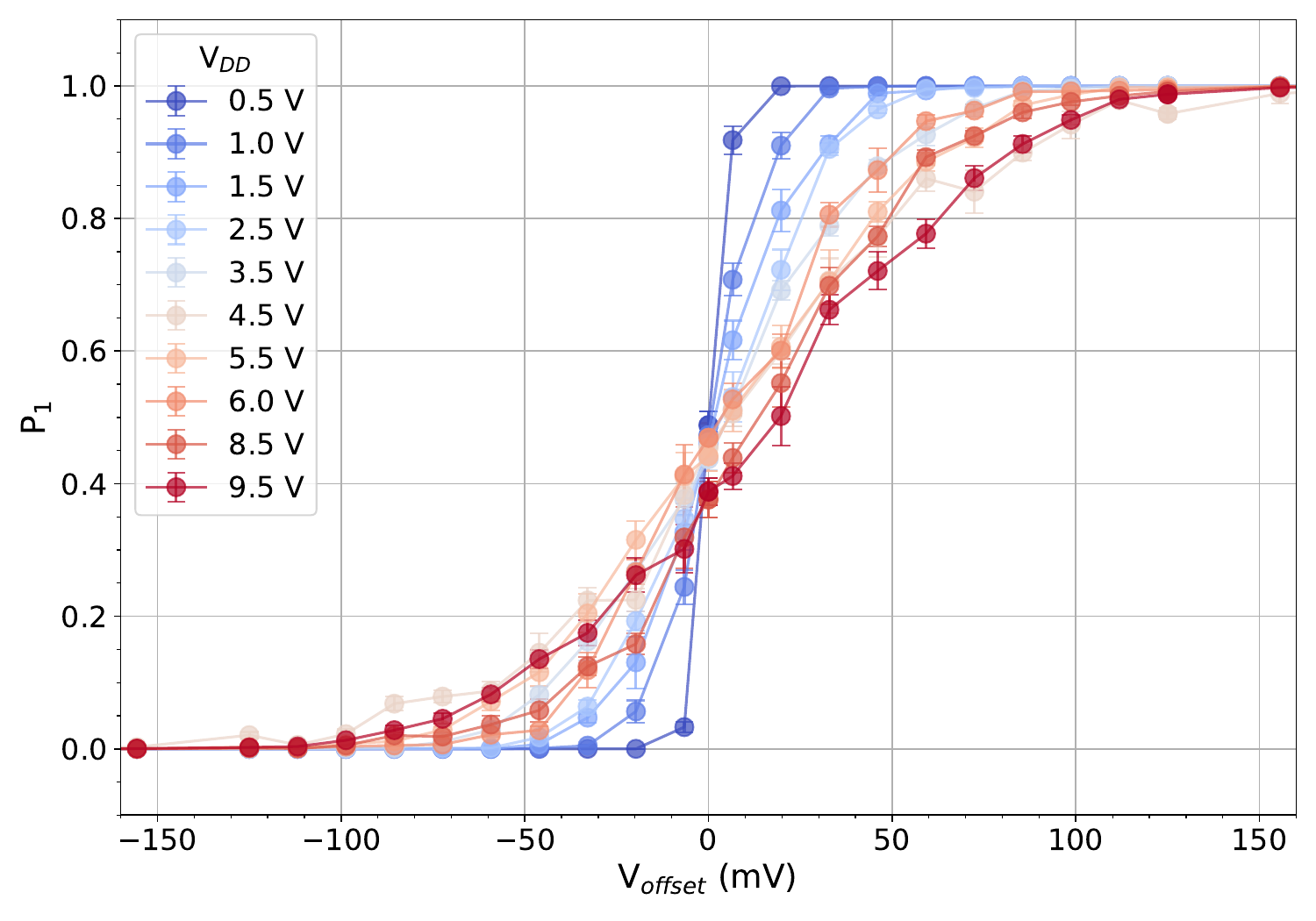}
\caption{\footnotesize{Ensemble average and standard deviation of \( P_1 \) at various offset voltages for multiple \( V_{DD} \) values. Each curve represents a different \( V_{DD} \), with error bars indicating the standard deviation at each offset voltage
($V_{\text{offset}}$). The fitted lines illustrate the expected binary stochastic neuron response (cf.\ {\bf{Eq.\ \ref{eq:stochastic_binary_neuron}}}). Ensemble statistics at each offset voltage are based on 5000 measurements.}}
\label{fig:fit}
\end{figure}

Building on prior theoretical formulations of stochastic binary neurons\cite{Camsari2017} 
and experimental demonstrations of sigmoidal transfer in hardware p-bits\cite{borders2019}, 
we model the probability $P_{1,i}$ of obtaining a logical ``1'' at the comparator output as a logistic function of the applied offset:
\begin{equation}
P_{1,i} =  
\left( \frac{1}{1 + \exp\!\left[-k \cdot \big(V_{\text{offset},i} - V_{\circ}\big)\right]} \right),
\label{eq:stochastic_binary_neuron}
\end{equation}
where $V_{\text{offset},i}$ is the applied bias offset, $k$ is the steepness factor of the response, 
and $V_{\circ}$ is the effective offset point determined primarily by the comparator’s intrinsic differential voltage. 
Experimentally, $P_{1,i}$ is extracted from the statistics of the TTL output signal in the circuit for the p-bit primitive 
(cf.\ {\bf{Figure \ref{fig:circuit}}}), where 
$V_{\text{out},i}$ fluctuates between $\sim$0~V and $V_{logic}=5$~V, 
and each instance with $V_{\text{out},i} \geq 2.5$~V is registered as a logical ``1.'' 
The fraction of such events over many trials corresponds to $P_{1,i}$, consistent with the 
stochastic neuron framework established in prior work\cite{Camsari2017,borders2019}.

\textbf{Figure~\ref{fig:fit}} presents the transfer response of the circuit in \textbf{Figure~\ref{fig:circuit}} when the memristive element $R_{\text{mem}}$ is realized with a two-terminal \activepoly\ memristor and the driving voltage $V_{DD}$ is swept from 0.5~V to 9.5~V. 
Each data point represents a sampling population of 5000 measurements and reports the mean value with the corresponding standard deviation for the given $V_{DD}$,$V_{\text{offset}}$ pair.
The form of the transfer curve is strongly dependent on $V_{DD}$; at lower values, the transition is steep and closely approximates an ideal probabilistic sigmoid with sharp thresholding, while at higher $V_{DD}$ the response becomes more gradual. 
This change reflects the scaling of $V_{\text{mvd}}$ at the divider, which directly modulates the steepness parameter $k$.  
The divider voltage noise increases linearly with the supply voltage, with its magnitude determined by the balance between the shunt conductance and the average memristor conductance, and the effect is strongest when the two conductances are comparable. As $V_{DD}$ increases, the divider amplifies memristor conductance fluctuations into larger voltage noise, broadening the stochastic response (cf.\ {\bf Eq.~\ref{eq:stochastic_binary_neuron}}) and reducing the slope parameter $k$. At lower $V_{DD}$, the input-referred noise is smaller, resulting in a sharper transition. The enhanced noise at higher $V_{DD}$ is reflected in the broader spread of the measured $P_1$ values, evident from the increasing standard deviations of their averages.

\begin{wrapfigure}{l}{85mm}
\centering
\includegraphics[trim={4mm 4mm 4mm 4mm}, clip, width=80mm]{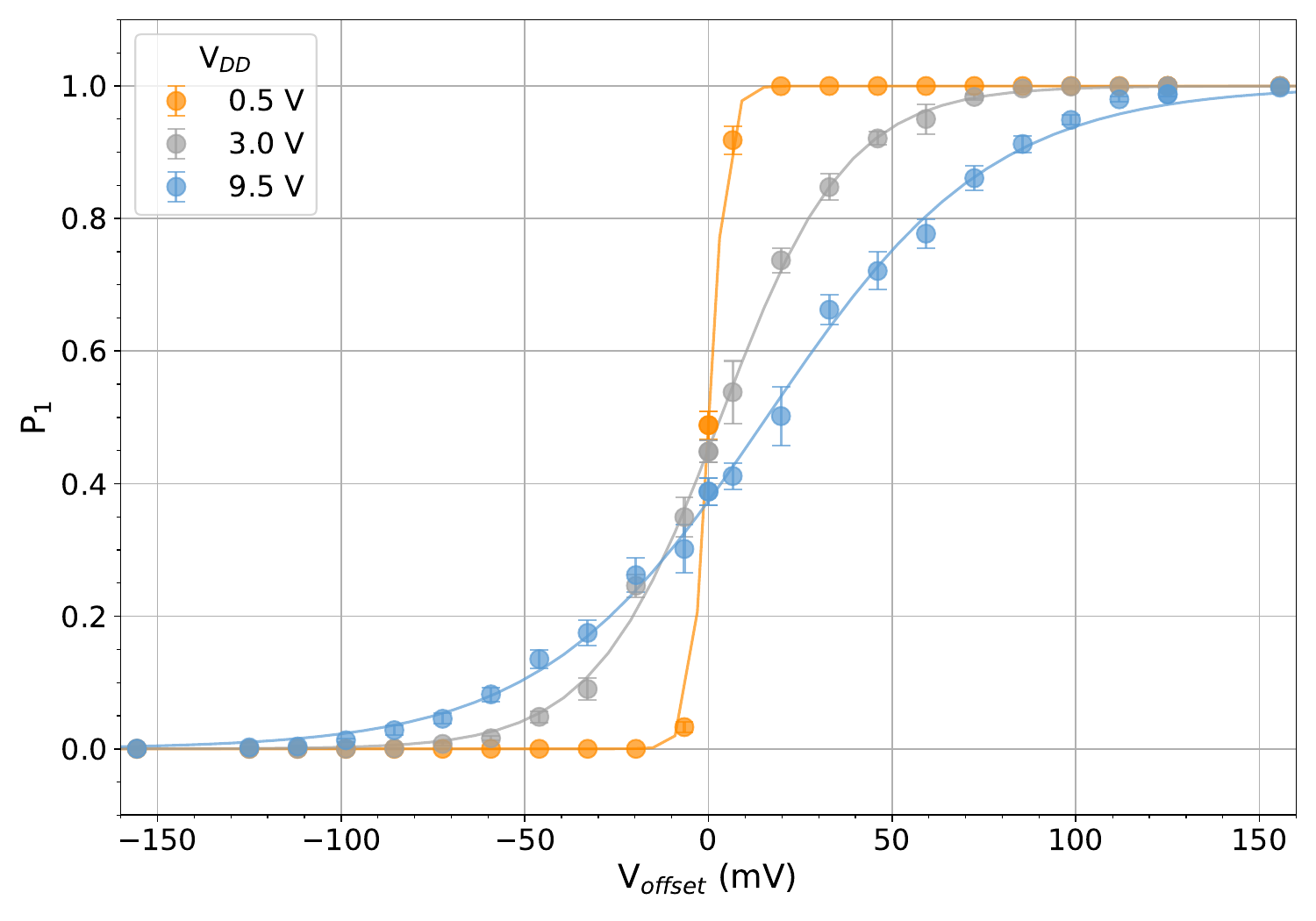}
\caption{\footnotesize 
Logistic fits (Eq.~\ref{eq:stochastic_binary_neuron}) to 
representative stochastic transfer responses at $V_{DD}=\,$0.5, 3.0, and 9.5~V. 
Reducing $V_{DD}$ sharpens the transition, increasing the slope near threshold, 
while increasing $V_{DD}$ broadens the response and shifts the effective threshold due to comparator offsets. 
Fit parameters are summarized in Table~\ref{tab:fit_params}.}
\label{fig:fitSel}
\end{wrapfigure}

\textbf{Figure~\ref{fig:fitSel}} compares the logistic fits of {\bf{Eq. \ref{eq:stochastic_binary_neuron}}} to representative experimental responses at $V_{DD}$=0.5, 3.0, and 9.5~V, while the extracted fitting parameters are listed in 
{\bf{Table~\ref{tab:fit_params}}}. As the supply is reduced from 9.5~V to 0.5~V, $k$ increases by nearly an order of magnitude, reflecting the sharper transition, while $V_{\circ}$ shifts systematically toward positive offset voltages. This rightward shift moves the 50\% crossing point away from zero bias, a consequence of comparator loading that skews the effective operating point.  

These results underscore that the stochastic transfer function is tunable through $V_{DD}$; 
lower supply voltages sharpen the response, increasing the slope near threshold but narrowing the effective range over which offsets influence the output, 
while higher voltages broaden the transition and distribute probabilities more gradually across $V_{\text{offset}}$. 
This tunability provides a direct means of modulating the spread of the output probability distribution, thereby linking circuit-level biasing to device-level stochastic behavior.
\begin{table}[h]
\centering
\begin{tabular}{c c c}
\toprule
$V_{DD}$ (V) & $k$ ((mV)$^{-1}$) & $V_{\circ}$ (mV) \\
\midrule
0.5 & 0.4229 & 0.1536 \\
3.0 & 0.0581 & 3.3205 \\
9.5 & 0.0325 & 15.8403 \\
\bottomrule
\end{tabular}
\caption{Logistic fit parameters extracted from Eq.~\ref{eq:stochastic_binary_neuron}. 
The steepness factor $k$ quantifies the sharpness of the probabilistic switching response, with larger values corresponding to steeper transfer curves. 
The offset parameter $V_{\circ}$ marks the effective threshold where $P_{1}=0.5$, shifting positively with $V_{DD}$ due to comparator loading and biasing effects.}
\label{tab:fit_params}
\end{table}

\begin{wrapfigure}{r}{90mm}
\vspace{-10pt}
\centering
\includegraphics[trim={4mm 4mm 4mm 4mm}, clip, width=85mm]{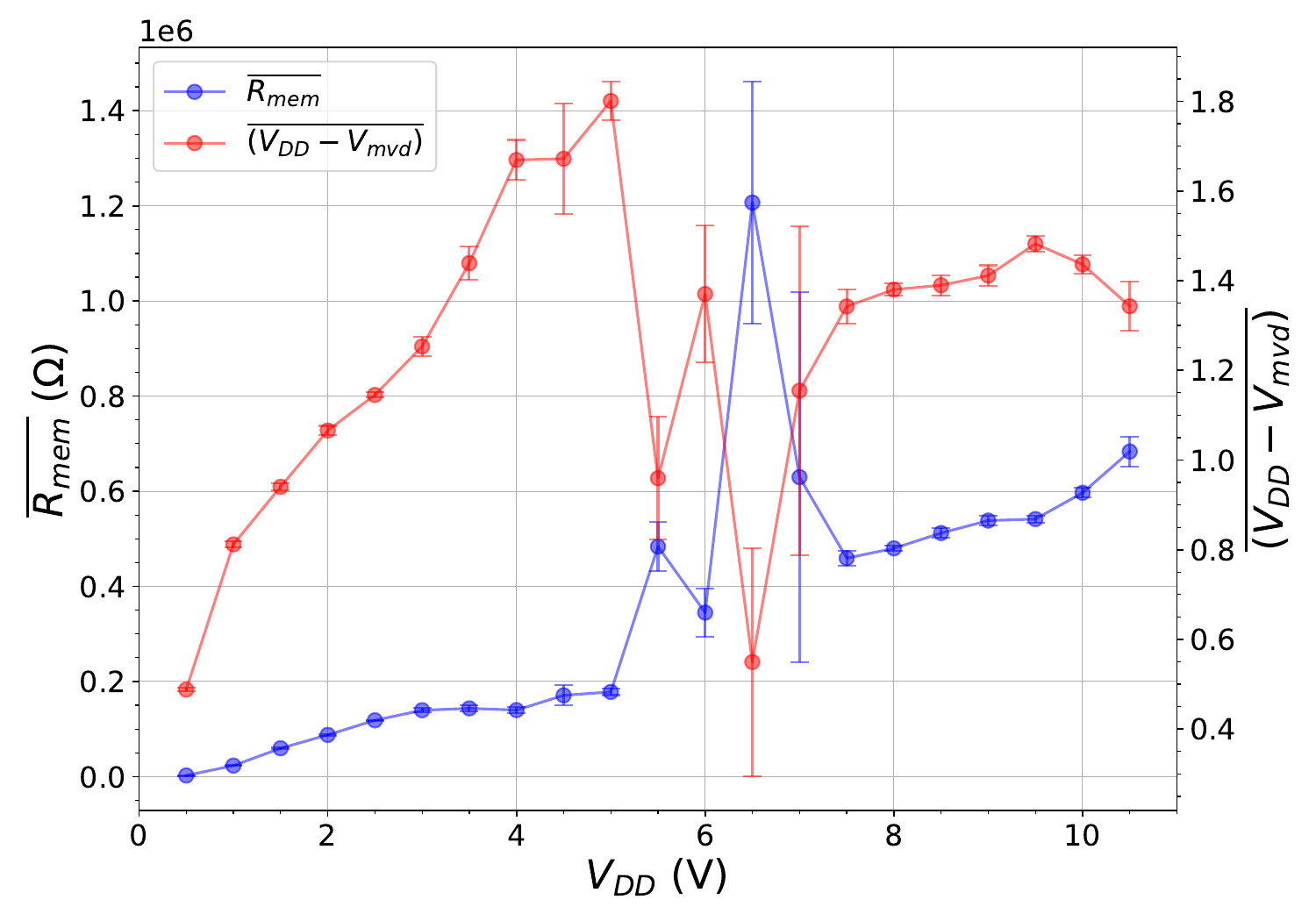}
\caption{\footnotesize{Average memristor resistance ($\overline{R}_{\text{mem}}$, blue) and 
voltage difference ($V_{DD}-V_{mvd}$, red) as functions of $V_{DD}$. 
Error bars denote standard deviations across measurement sets.}}
\label{fig:rmem}
\end{wrapfigure}

{\bf{Figure \ref{fig:rmem}}} presents a dual-axis plot of the average memristor resistance 
($\overline{R_{\text{mem}}}$) and the corresponding voltage drop across the device 
($\overline{V_{DD}-V_{mvd}}$) as functions of $V_{DD}$. Both quantities exhibit a pronounced 
disturbance in the 5–7~V range. Within this bias window, $\overline{R_{\text{mem}}}$ departs 
from its otherwise monotonic decrease, fluctuating between 0.8~M$\Omega$ and 1.4~M$\Omega$, 
while $\overline{V_{DD}-V_{mvd}}$ deviates from the linear progression evident at lower and 
higher voltages. 

These anomalies signal the onset of an unstable or transitional regime in which 
conformational rearrangements of the polymer backbone and pendant groups, 
charge accumulation, and redox-driven changes in percolation pathways couple 
directly to the voltage-divider and comparator dynamics. The interplay between 
intrinsic device variability and circuit loading produces nonlinear feedback that 
amplifies fluctuations in both $\overline{R_{\text{mem}}}$ and $V_{DD}-V_{mvd}$.
In this 
regime, the device does not exhibit the smooth scaling of resistance with applied bias that is 
expected in a stable memristive state, but instead samples multiple metastable configurations.

\begin{wrapfigure}{r}{90mm}
\vspace{-10pt}
\centering
\includegraphics[trim={2mm 2mm 2mm 2mm}, clip, width=85mm]{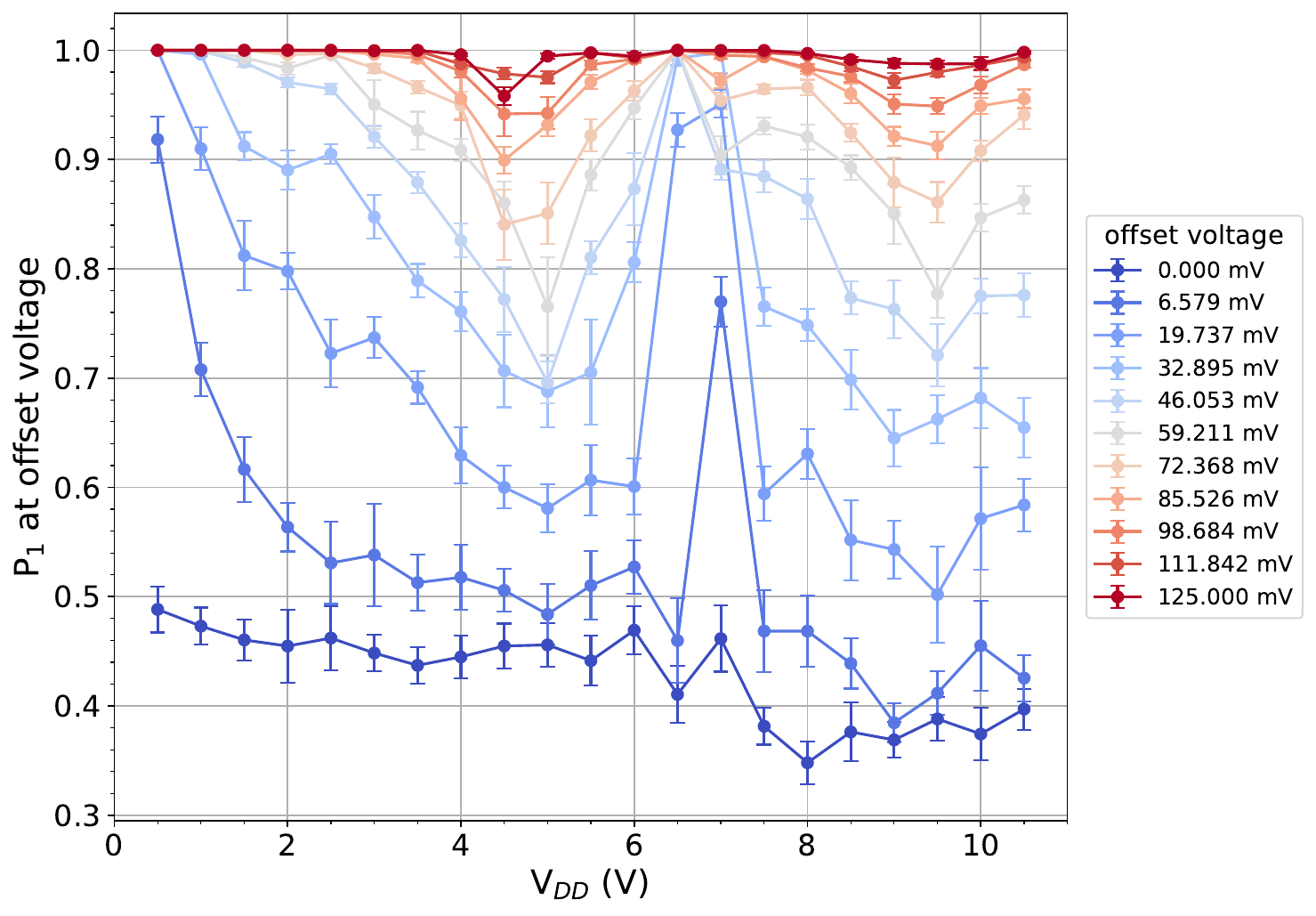}%
\caption{\footnotesize{Probability $P_{1}$ as a function of $V_{DD}$ for a series of measured offset voltages (0--125~mV).}}
\label{fig:chaos01}
\end{wrapfigure}

To evaluate how this disturbance propagates into circuit-level performance, we measured 
the probability $P_{1}$ as a function of $V_{DD}$ at fixed offset voltages. {\bf{Figure \ref{fig:chaos01}}} 
summarizes these results for offsets spanning 0–125~mV. Each curve traces the evolution 
of $P_{1}$ with increasing $V_{DD}$ under constant offset bias. At low offsets, $P_{1}$ remains 
near 0.5 across much of the voltage range, reflecting a balanced stochastic output. As the 
offset increases, however, the curves shift systematically upward, demonstrating the ability 
of even modest biases to skew the probability distribution toward unity. Importantly, the 
steepest portions of these curves occur within the same 5–7~V disturbance window identified 
in {\bf{Figure \ref{fig:rmem}}}. This correlation indicates that the transition-state variability of 
the memristive network is directly imprinted onto the probabilistic transfer characteristics 
of the circuit.  

The combined results emphasize two key points. First, the disturbance regime is not 
a benign fluctuation but a manifestation of the network exploring competing conductive 
pathways, producing irregular and bias-sensitive responses. Second, this microscopic 
instability translates into macroscopic unpredictability in $P_{1}$, where small changes in 
$V_{DD}$ can cause disproportionately large shifts in the probability distribution. For p-bit 
operation, this duality is both a challenge and an opportunity. The reduced stability 
observed in this bias window poses challenges for reproducibility. 
At the same time, the results highlight the direct connection between 
conformational fluctuations within the polymer network and the statistical properties of 
the p-bit. This regime therefore represents a bias-controlled domain where device-level 
dynamics govern the accessible probability distributions, providing a mechanism to 
modulate entropy generation at the circuit level. In this sense, {\bf{Figures \ref{fig:rmem}}} 
and {\bf{\ref{fig:chaos01}}} together show that controlled exploitation of the disturbance 
region can enable dynamic adjustment of entropy output, though practical implementations 
must carefully balance sensitivity against stability to harness this behavior for 
probabilistic computing.  

\begin{figure}[h]
\begin{center}
\scalebox{0.45}[0.45]{\includegraphics{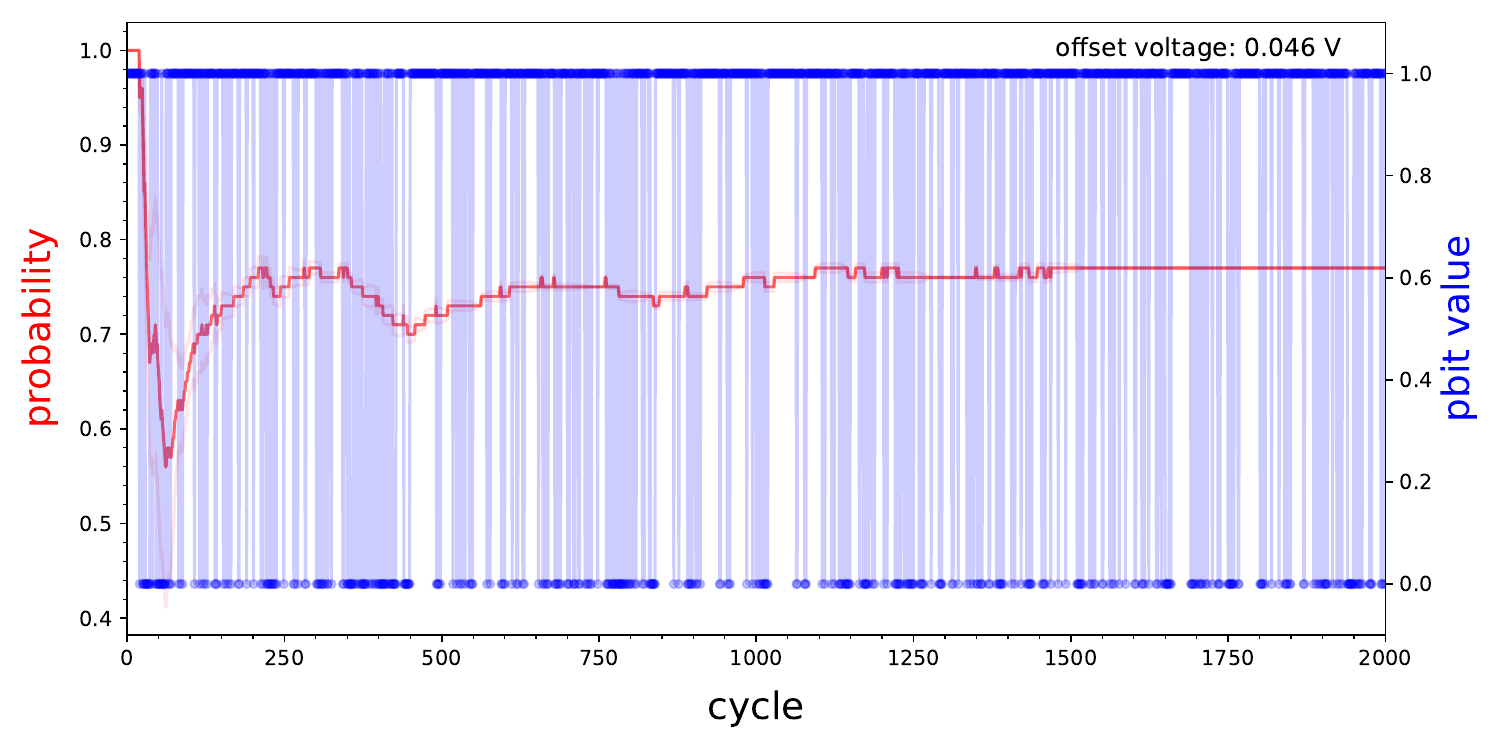}}
\caption{\footnotesize{Probabilistic bit \(P_1\) behavior across the first 2000
measurement cycles with offset voltage of 
 \(46\) mV at  V$_{DD}$=9.0 V.
The red line represents the running average of \(P_1\).
}}
\label{fig:pbit_046}
\end{center}
\end{figure}

Building on this perspective, {\bf{Figure~\ref{fig:pbit_046}}} illustrates the stochastic 
switching dynamics of a single p-bit at a fixed offset of \SI{46}{\milli\volt} and 
$V_{DD}=\SI{9.0}{\volt}$. While individual cycles fluctuate randomly, the running average 
of $P_{1}$ progressively stabilizes to $\sim 0.7$--$0.8$, converging after approximately 
1500 samples. This example demonstrates how even modest offsets bias the probability 
distribution away from the symmetric case of $P_{1}=0.5$, yet preserve randomness at 
the single-bit level. The result directly connects the ensemble-level variability discussed 
above with time-resolved convergence behavior, underscoring voltage-controlled tunability 
as a defining feature of p-bit operation.

\subsection*{Memristive device response}

The stochastic transfer characteristics previously described ultimately originate from the intrinsic memristive dynamics of the active polymer element.
Since their theoretical proposal in the early 1970s, memristors have been associated with several characteristic electrical signatures that facilitate their identification. One hallmark is the presence of a \textit{pinched hysteresis loop} in the current--voltage (\(I\!-\!V\)) response under an applied AC voltage, which narrows progressively as the driving frequency increases\cite{lc:pinched01}. 

\begin{wrapfigure}{r}{85mm}
\centering
\includegraphics[trim={4mm 4mm 4mm 4mm}, clip, width=80mm]{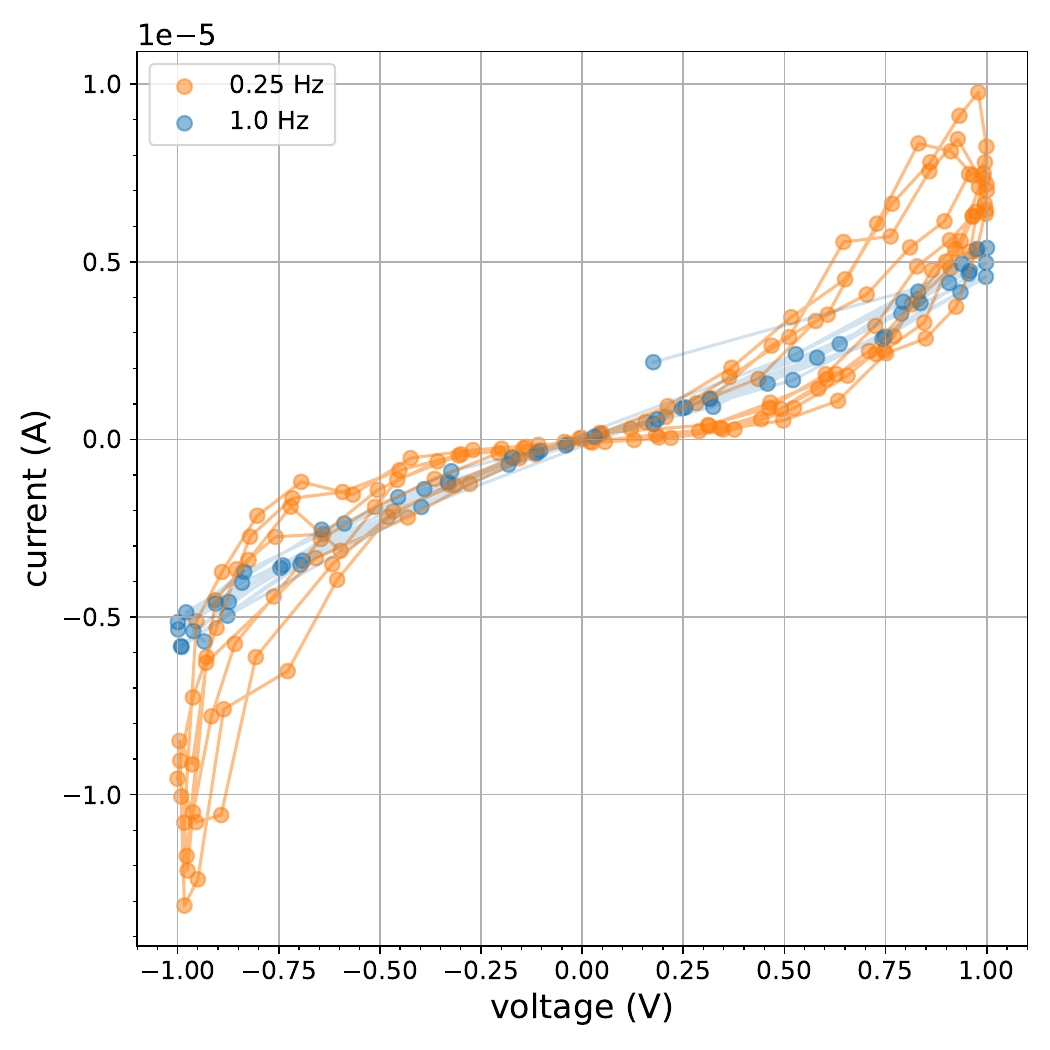}
\caption{\footnotesize Frequency-dependent AC \(I\!-\!V\) characteristics of two-terminal ITO/\activepoly/Al devices showing pinched hysteresis that diminishes with increasing frequency, consistent with memristive behavior. Measurements taken at 23 \degC.}
\label{fig:memistor}
\end{wrapfigure}
\textbf{Figure~\ref{fig:memistor}} presents the AC \(I\!-\!V\) response of a \activepoly\ memristor  under a \(\pm1~\mathrm{V}\) sinusoidal excitation. At \SI{0.25}{\hertz}, the device exhibits a pronounced pinched hysteresis loop, consistent with memristive behavior, which decreases in area as the frequency increases to \SI{1}{\hertz}. Specifically, the loop area \(\oint I\,\mathrm{d}V\) decreases from \num{3.41e-6}~A$\cdot$V at \SI{0.25}{\hertz} to \num{4.26e-7}~A$\cdot$V at \SI{1.0}{\hertz}. Concurrently, the linearity of the \(I\!-\!V\) response, captured by the coefficient of determination (\(R^2\)) from a linear fit, increases from 0.851 to 0.981. This transition from a pinched hysteresis loop to a linear resistive profile reflects the diminishing memory response under rapid voltage modulation and confirms the dynamic, frequency-sensitive behavior of the \activepoly\ devices.

Dielectric analysis was used to probe the molecular motions that underlie this frequency dependence and revealed two thermally activated processes. A low-temperature relaxation with \(E_{a}\approx 0.44~\text{eV}\) and \(f_{0}\approx 6\times 10^{7}~\text{Hz}\) is consistent with \(\beta\)-type motions, such as ether-linkage and hexyl-segment rotations or phenyl flips on the N,N-diphenylaniline (TPA) headgroup. At room temperature (\(T=296~\text{K}\)), this barrier yields \(E_{a}/k_{B}T \approx 17.3\).
Inserting this into \(f=f_{0}\exp(-E_{a}/k_{B}T)\) yields \(f_{0}\approx 3.1\times 10^{7}~\text{Hz}\) at 1~Hz and \(f_{0}\approx 7.8\times 10^{6}~\text{Hz}\) at 0.25~Hz, 
yielding characteristic frequencies in the 1--2~Hz range, which are of the same order 
as the observed transition between 0.25 and 1~Hz, indicating reasonable consistency 
with the dielectric fit given uncertainties in $f_{0}$ and circuit loading effects.
A higher-temperature process with \(E_{a}\approx 1.04~\text{eV}\) likely reflects cooperative backbone or segmental rearrangements, although the absence of a distinct calorimetric glass transition between \(-90~^{\circ}\text{C}\) and \(200~^{\circ}\text{C}\) suggests that any \(\alpha\)-relaxation is shallow or mixed with other modes. 
\begin{wrapfigure}{r}{90mm}
\centering
\includegraphics[trim={12mm 4mm 16mm 12mm}, clip, width=80mm]{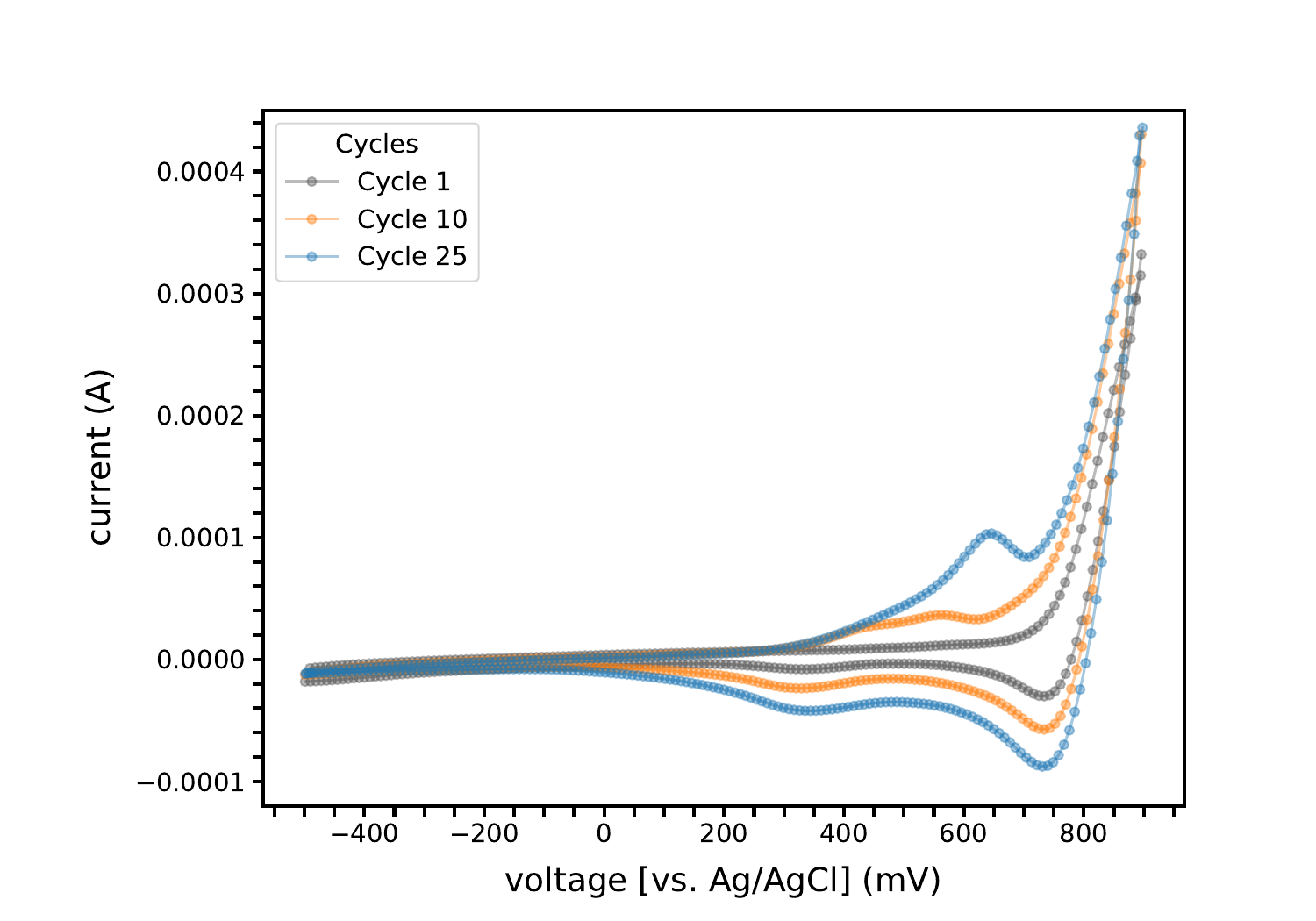}
\caption{\footnotesize Electropolymerization CVs (1st, 10th, 25th cycles) of \activepoly\ on ITO in ACN/0.1~M TBAPF$_6$; scan \SI{100}{mV.s^{-1}}; \SI{-0.5}{V} to \SI{+0.9}{V} vs Ag/AgCl.}
\label{fig:yb1766_electro}
\end{wrapfigure}
Within this framework, resistive switching is governed by shallow conformational potentials and localized carrier dynamics, where reorientation of triphenylamine side groups or short-range hopping events define the frequency dependence of loop closure and couple intrinsic polymer motions to the stochastic resistive fluctuations central to p-bit operation. As the excitation frequency increases, the voltage cycle becomes shorter than the relaxation time of these motions, limiting their ability to reorient within a single cycle. This suppression narrows the hysteresis and produces a transition from memristive to more resistive-like behavior. Prior studies on DTP polymers with pendant carbazole groups indicated a multipercolation mechanism, where intermittent side-group reorientations modulated conductive pathways and current response~\cite{foulger2021}. A comparable mechanism is likely active here, with pendant triphenylamines dynamically reshaping the percolation network and driving the observed memristive behavior.

\subsection{Device formation and operating window}

To translate the molecular relaxation processes identified earlier into a measurable 
device response, we constructed two-terminal memristive elements using \activepoly\ 
as the active layer. These devices, employed in the circuit of 
\textbf{Figure~\ref{fig:circuit}}, provide the platform where conformational dynamics 
of the polymer couple to circuit-level stochasticity. 
For reliable evaluation of this 
behavior, uniform and reproducible thin films are essential. To this end, 
\activepoly\ was electropolymerized directly onto ITO substrates, and 
\textbf{Figure~\ref{fig:yb1766_electro}} presents representative cyclic voltammograms 
acquired during the electropolymerization process.

\begin{wrapfigure}{r}{105mm}
\centering
\includegraphics[trim={4mm 4mm 4mm 4mm}, clip, width=100mm]{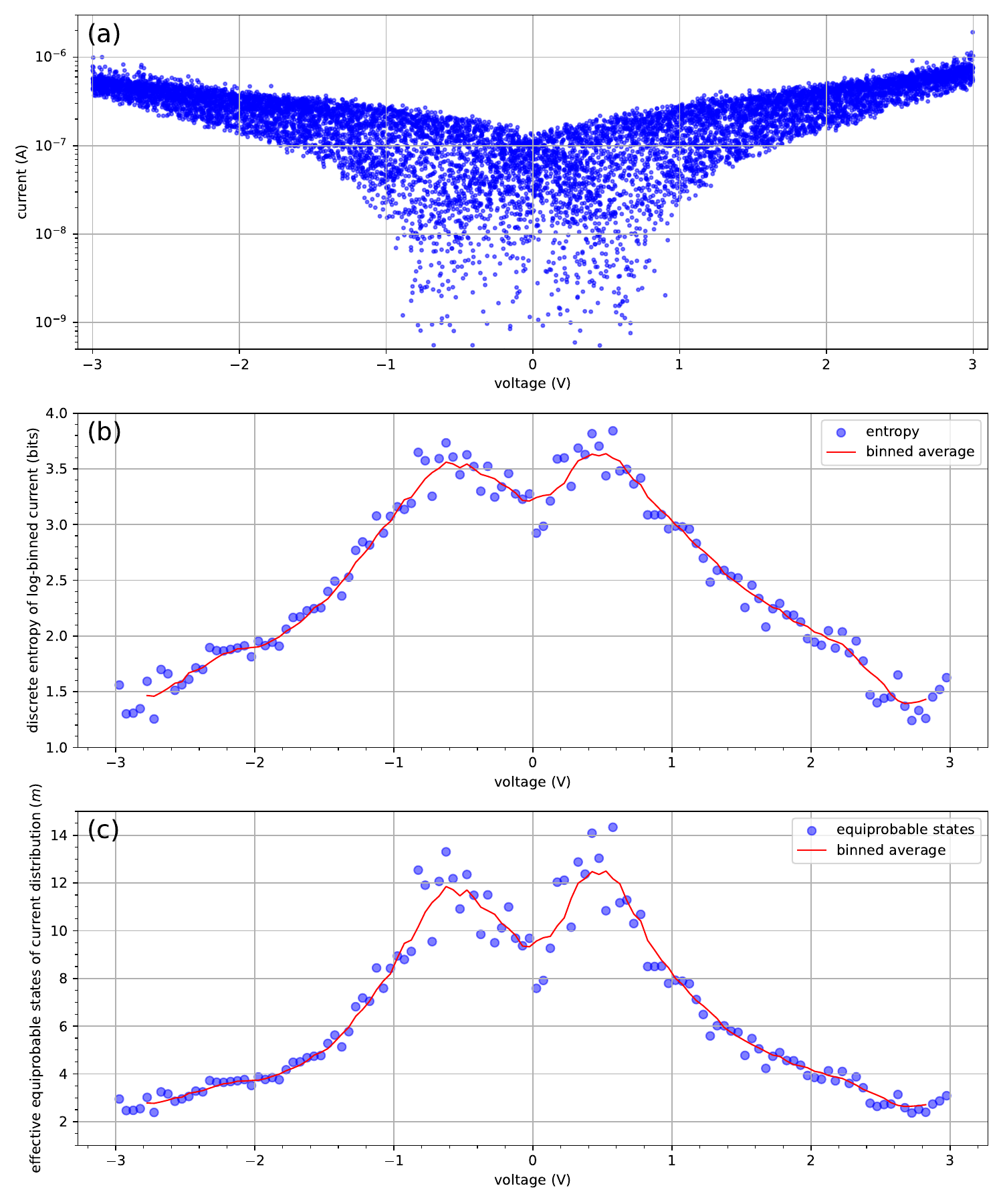}%
\caption{\footnotesize{Voltage dependence of information metrics for ITO/\activepoly/Al devices derived from DC $I$--$V$ data: 
\textbf{(a)} Scatter of current under randomized DC voltage application; 
\textbf{(b)} discrete Shannon entropy $H_{\mathrm{disc}}$ (bits) of the log-binned analog current distributions;
\textbf{(c)} corresponding effective number of equiprobable states, 
$N_{\mathrm{eff}} = 2^{H_{\mathrm{disc}}}$.}}
\label{fig:entropy}
\end{wrapfigure}

Polymerization was carried out by cycling the potential between --500 mV and +900 mV versus Ag/AgCl in an acetonitrile (ACN) solution containing 0.1~M TBAPF\textsubscript{6} and the monomer. 
A total of 25 cycles were applied at a scan rate of 100~mV/s. 
Representative voltammograms from the 1st, 10th, and 25th cycles are presented
(cf.\ \textbf{Figure~\ref{fig:yb1766_electro}}). 
With increasing cycle number, both oxidative and reductive currents grow in magnitude, reflecting continuous film growth and an expansion of the electroactive surface area. 
Progressive sharpening of the redox peaks and increased current density are indicative of uniform film growth and enhanced charge carrier mobility.
Visually, the resulting films exhibit a deep blue coloration upon oxidation, characteristic of the polymer’s fully doped state.
Following electropolymerization, the films were subjected to a final conditioning step at a fixed potential of +900~mV for 30~s in monomer-free electrolyte to ensure complete oxidation prior to Al evaporation.
Cyclic voltammetry of the fully formed film in a monomer-free electrolyte revealed an oxidation onset corresponding to a HOMO energy level of 
$-4.8~\mathrm{eV}$. 
When combined with the optical bandgap of 
$2.83~\mathrm{eV}$ (438~nm) obtained from UV-Vis absorption, the LUMO energy was estimated to be 
$-2.0~\mathrm{eV}$.
Finally, an Al top contact was then deposited over the polymer, yielding an ITO/\activepoly/Al device structure.

\subsection{Stochastic DC Response and Variability at Fixed Bias}

A key property enabling the use of \activepoly\ devices as fluctuation-resistive elements in p-bit circuits is the variability in current when the device is repeatedly biased at the same voltage. This temporal variability originates from field- and thermally driven reconfiguration of the percolation network formed by the DTP backbone and pendant TPA units. While not itself proof of memristance, this stochastic response is central to voltage-gated probabilistic operation. The memristive character is established independently by the frequency-dependent pinched hysteresis loops presented in {\bf Figure~\ref{fig:memistor}}.

{\bf Figure~\ref{fig:entropy}a} presents the square-wave pulsed DC $I$--$V$ response, highlighting the nonlinear and temporally variable current output under fixed bias. Current levels span over three orders of magnitude across the $\pm 3$~V range, and even within the $\pm 1$~V operating window relevant for p-bit function 
(cf.\ {\bf Figure~\ref{fig:rmem}})
the device samples multiple current states at each bias. 
To quantify this variability, the ensemble of currents collected within each voltage bin was converted into a normalized probability distribution $P(I\mid V)$ using logarithmically spaced current bins. From these distributions, the discrete Shannon entropy 
\begin{equation}
H_{\mathrm{disc}}(I \mid V) = - \sum_i P(I_i \mid V)\,\log_{2} P(I_i \mid V).
\label{eq:disc_entropy}
\end{equation}
was computed, providing a measure of the information content of the current response (\textbf{Figure~\ref{fig:entropy}b}). The corresponding effective number of equiprobable states, $N_{\mathrm{eff}}(I \mid V) = 2^{H_{\mathrm{disc}}(I \mid V)}$, is shown in \textbf{Figure~\ref{fig:entropy}c} as an intuitive representation of the breadth of accessible current outcomes.

Voltage ranges with elevated $H_{\mathrm{disc}}$ and correspondingly large $N_{\mathrm{eff}}$ indicate conditions where the memristor explores a wide spectrum of current configurations, a prerequisite for robust p-bit operation. A distinct maximum in entropy emerges within a finite bias window, signaling a regime where multiple conduction channels contribute with near-comparable probability, thereby maximizing configuration-space sampling. Importantly, this entropy maximum coincides with the 0.5--0.8~V range of $\overline{V_{DD}-V_{\mathrm{mvd}}}$ fluctuations identified in {\bf Figure~\ref{fig:rmem}}, where the device exhibits the strongest variability in its transfer characteristics.
Although entropy is derived from ensembles of pulsed DC measurements 
(cf.\ \textbf{Figure~\ref{fig:entropy}a})
and the transfer function from the p-bit primitive circuit (cf.\ \textbf{Figure~\ref{fig:circuit}}), 
both converge on the same physical origin, the nonlinear conformational response of the polymer under bias. 
This correspondence demonstrates that the voltage window producing peak entropy is also the regime of maximal stochasticity in circuit behavior. These conditions highlight the bias range where p-bit functionality is both most effective and most tunable.

\subsection{Energy-resolved electrochemical impedance spectroscopy}

\begin{figure}[t]
\centering
\subcaptionbox{}[0.9\textwidth]{%
    \includegraphics[trim={6mm 3mm 4mm 6mm}, clip, width=90mm]{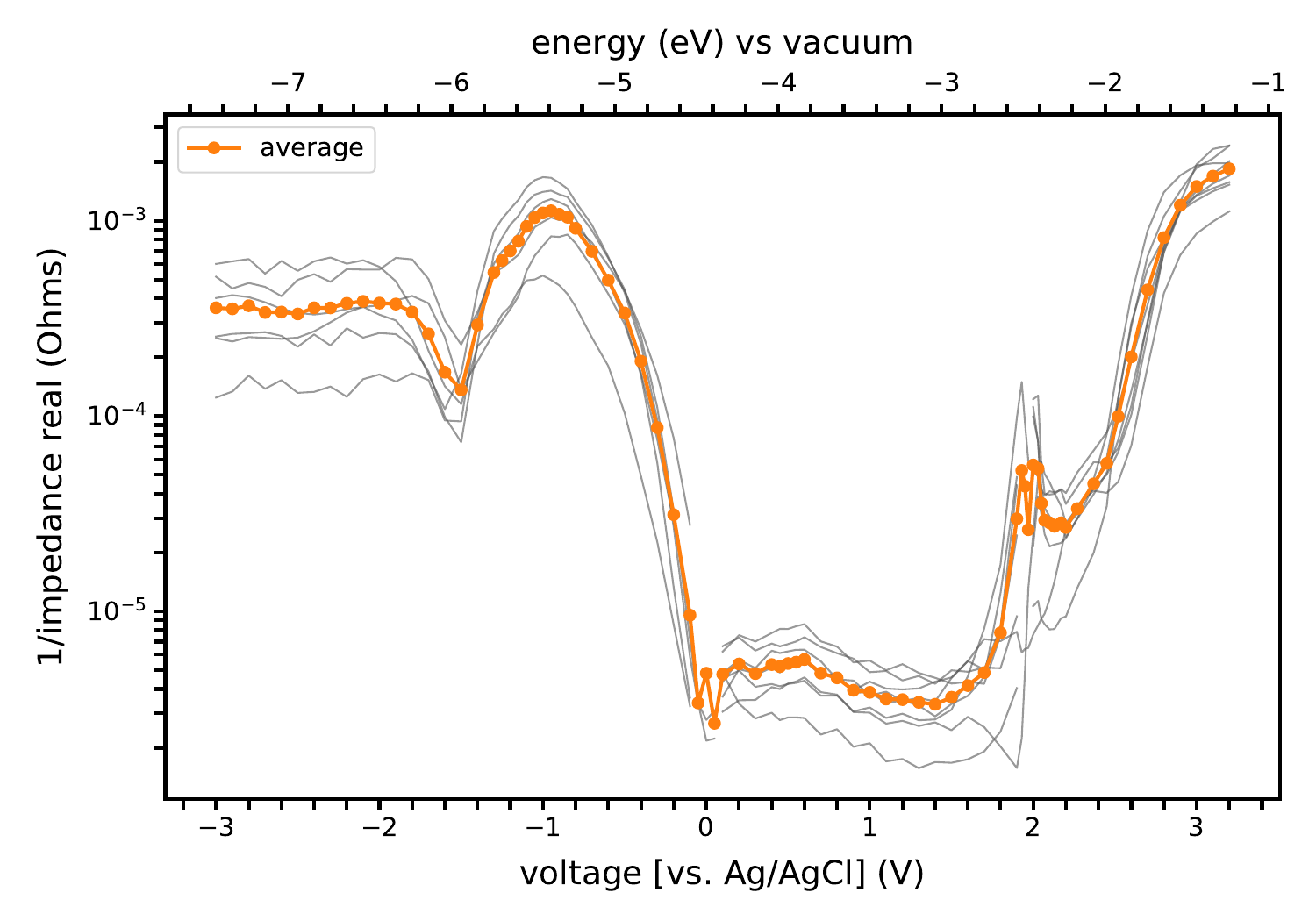}%
}
\subcaptionbox{}[0.9\textwidth]{%
    \includegraphics[trim={6mm 3mm 4mm 6mm}, clip, width=90mm]{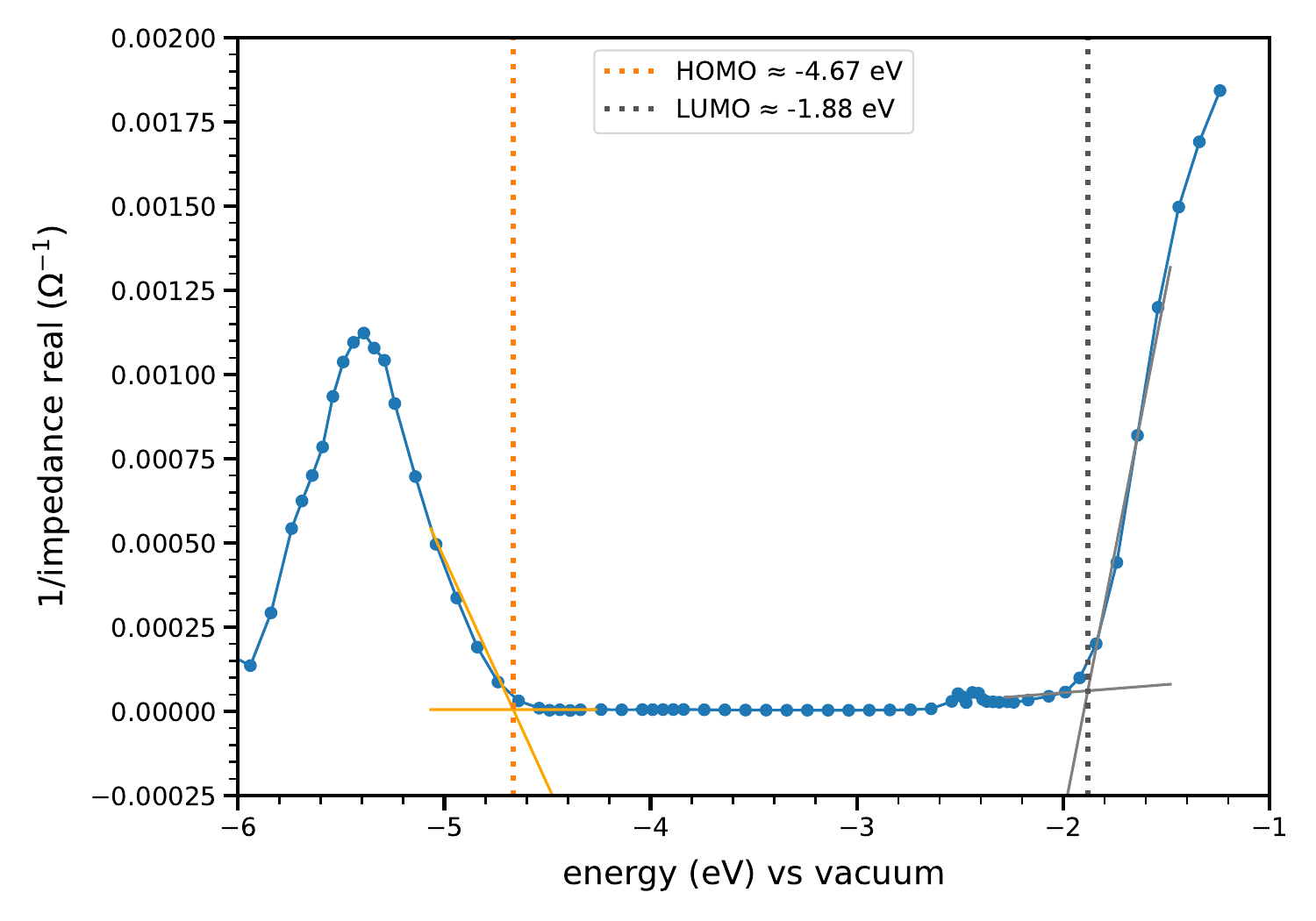}%
}
\caption{ER--EIS characterization of \activepoly\ films: (a) average (orange) of six runs (grey), HOMO \SI{-4.67}{eV}, LUMO \SI{-1.88}{eV}; (b) linear-scale onset extraction showing a transport gap of $\sim$\SI{2.79}{eV}.}
\label{fig:yb1766_electro_HOMO-LUMO}
\end{figure}

The bias-localized entropy maximum identified above points to an electronic origin rooted 
in the structure of \activepoly. To probe this connection experimentally, we employed 
energy-resolved electrochemical impedance spectroscopy (ER-EIS)~\cite{EREIS_2014,schauer2018} 
to directly map the density of states (DOS) of electropolymerized films. As shown in 
\textbf{Figure~\ref{fig:yb1766_electro_HOMO-LUMO}a}, six independent spectra and their 
average exhibit excellent consistency, with onset variations of less than 
$\pm0.02~\mathrm{eV}$. The averaged DOS reveals sharp, symmetric rises at the HOMO and 
LUMO edges with minimal mid-gap noise, while slight curvature ahead of each onset suggests 
shallow localized states typical of disordered organic semiconductors. Linear-scale plots 
(cf.\ \textbf{Figure~\ref{fig:yb1766_electro_HOMO-LUMO}b}) enable precise extrapolation of 
band edges~\cite{EREIS_2015}, yielding HOMO and LUMO onsets of $-4.67~\mathrm{eV}$ and 
$-1.88~\mathrm{eV}$ (vs.\ vacuum). The resulting $2.79~\mathrm{eV}$ transport gap agrees 
closely with the $2.83~\mathrm{eV}$ gap derived from UV–Vis spectroscopy.  

The HOMO and LUMO energy levels obtained from cyclic voltammetry, UV--visible absorption, 
and ER--EIS are summarized in {\bf{Table~\ref{table:bandgap_comparison}}} and show excellent 
consistency with previous reports for \activepoly~\cite{foulger:2025_01}. Their close 
alignment with the ITO work function ($\sim -4.7~\mathrm{eV}$) supports efficient hole 
injection and oxidation-driven charge transport in the device. While these experimental 
methods establish the energetic landscape at the film level, they do not resolve how the 
polymer’s molecular architecture gives rise to the observed band alignment or the bias-localized 
stochasticity. To address this, we performed density functional theory (DFT) calculations 
on oligomeric segments of \activepoly\ to gain orbital-level insight into the distribution of 
frontier states and their evolution with oxidation state.  

\subsection{Electronic-structure mechanism (TDOS/PDOS/OPDOS)}

\begin{table}[hhh]
\begin{center}
\begin{tabular}{lccc}
\hline
 & HOMO (eV) & LUMO (eV) & bandgap (eV) \\
\hline \hline
cyclic voltammetry (CV)     & $-4.80$ & $-2.00$ & $2.83$ \\
ER--EIS                     & $-4.67$ & $-1.88$ & $2.79$ \\
DFT, 5-mer                  & $-4.27$ & $-2.19$ & $2.08$ \\
DFT, 2--8 mer range         & $-4.45 \rightarrow -3.99$ & $-1.36 \rightarrow -2.00$ & -- \\
\hline
\end{tabular}
\end{center}
\caption{Comparison of HOMO, LUMO, and bandgap energies obtained by three methods.}
\label{table:bandgap_comparison}
\end{table}

Density functional theory (DFT) calculations were performed on \activepoly\ oligomers 
(dimer through octamer) in an antiperi\-planar conformation. As the chain length increases 
from two to eight repeat units, the HOMO levels increase from $-4.45~\mathrm{eV}$ to 
$-3.99~\mathrm{eV}$, while the LUMO levels decrease from $-1.36~\mathrm{eV}$ to 
$-2.00~\mathrm{eV}$.  

\begin{figure*}[t]
\centering
\includegraphics[trim={40mm 25mm 40mm 50mm}, clip, width=0.95\textwidth]{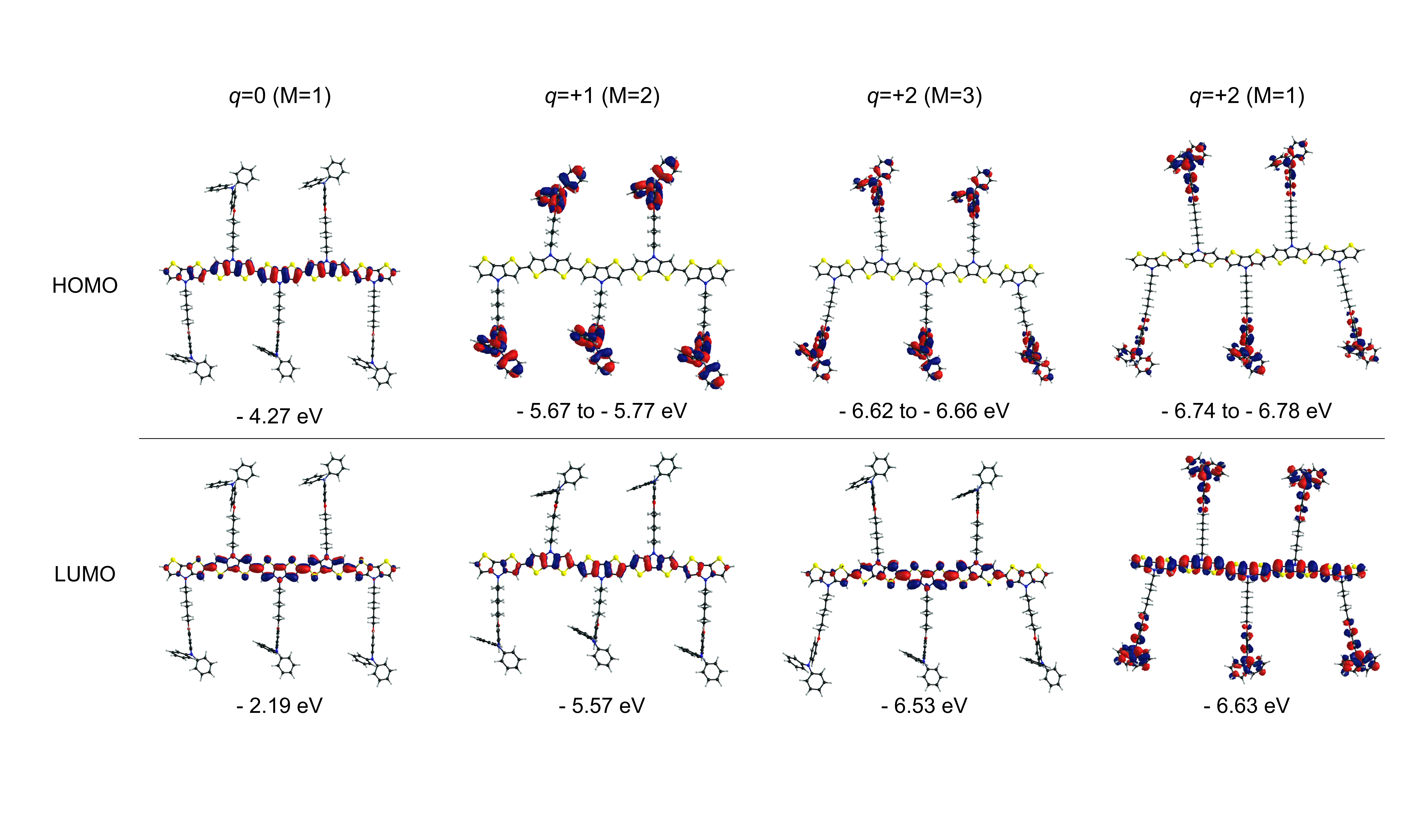}
\caption{\footnotesize 
DFT calculations of \activepoly\ oligomers in an antiperi\-planar 
conformation (B3LYP-D3BJ/def2-TZVP, gas phase). 
\textbf{Neutral pentamer ($q=0$):} The HOMO ($-4.27~\mathrm{eV}$) and LUMO ($-2.19~\mathrm{eV}$) 
localize on the DTP backbone, while deeper TPA-based orbitals ($-4.97$ to $-4.99~\mathrm{eV}$) 
remain unmixed, suggesting potential secondary hole-hopping channels upon oxidation. 
\textbf{Radical cation ($q=+1$):} The adiabatic ionization potential is $4.93~\mathrm{eV}$ and 
the $\beta$-spin HOMO--LUMO gap collapses to $\approx 0.10~\mathrm{eV}$; the $\beta$-HOMO is 
degenerate over all TPA units ($-5.67$ to $-5.77~\mathrm{eV}$), while the $\beta$-LUMO localizes 
on DTP ($-5.57~\mathrm{eV}$). 
\textbf{Dication diradical ($q=+2$):} 
In the triplet ($S=1$), $\alpha$- and $\beta$-HOMOs 
(HOMO to HOMO--4) are degenerate and delocalized over all TPA units ($E=-6.62$ to 
$-6.66~\mathrm{eV}$), while LUMOs localize exclusively on DTP ($E=-6.53~\mathrm{eV}$), yielding 
a $\beta$--$\beta$ gap of $0.09~\mathrm{eV}$. In the open-shell singlet ($S=0$), HOMOs reside 
on TPA ($E=-6.74$ to $-6.78~\mathrm{eV}$), and the LUMO ($E=-6.63~\mathrm{eV}$) is mainly on DTP 
with partial TPA character, giving a gap of $0.11~\mathrm{eV}$.  
}
\label{fig:yb1766_atomic}
\end{figure*}

For the neutral pentamer, both the HOMO ($-4.27~\mathrm{eV}$) and the LUMO ($-2.19~\mathrm{eV}$) 
localize on the DTP core, whereas triphenylamine (TPA) based orbitals lie slightly deeper 
($-4.97$ to $-4.99~\mathrm{eV}$) and do not mix with the HOMO, suggesting they could serve as 
secondary hole-hopping channels upon oxidation. {\bf{Figure~\ref{fig:yb1766_atomic}}} 
presents a visualization of the corresponding HOMO and LUMO orbitals. In the mono\-oxidized 
radical-cation (doublet) state, the adiabatic ionization potential is $4.93~\mathrm{eV}$ and 
the $\beta$-spin HOMO--LUMO gap collapses to $\approx 0.10~\mathrm{eV}$; the $\beta$-HOMO is 
degenerate over all five TPA units ($-5.67$ to $-5.77~\mathrm{eV}$), while the $\beta$-LUMO 
localizes on the DTP backbone ($-5.57~\mathrm{eV}$).  

Oxidation to the dication yields an open-shell diradical with two accessible spin states. In the triplet ($S=1$), both $\alpha$- and $\beta$-spin HOMOs (HOMO to HOMO--4) are degenerate and fully delocalized over all five TPA units ($E=-6.62$ to $-6.66~\mathrm{eV}$), while both $\alpha$- and $\beta$-spin LUMOs localize exclusively on the DTP backbone ($E=-6.53~\mathrm{eV}$), giving a $\beta$--$\beta$ gap of $0.09~\mathrm{eV}$. 

\begin{figure*}[t]
\centering
    \includegraphics[trim={1mm 1mm 1mm 1mm}, clip, width=0.95\textwidth]{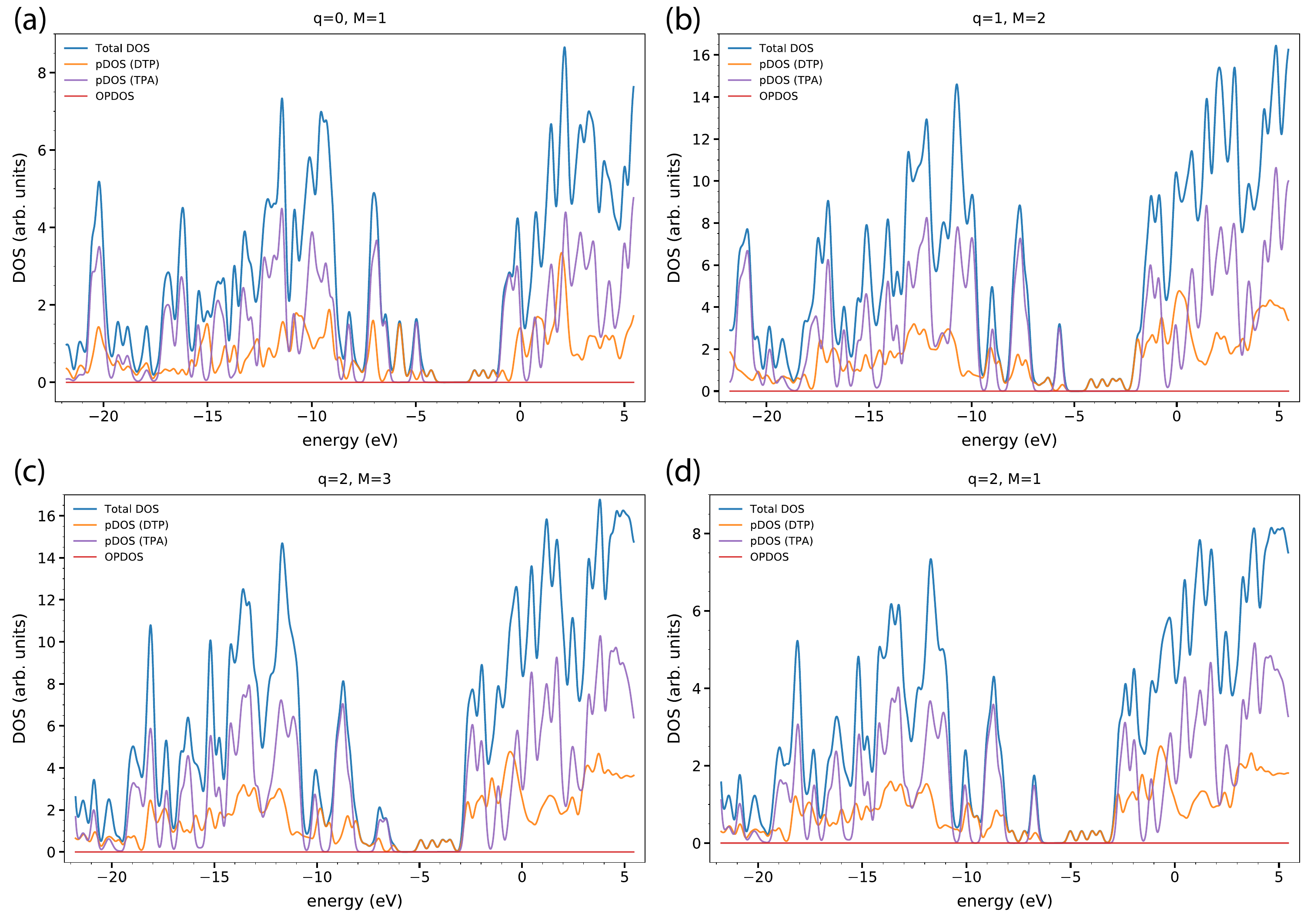}
\caption{\footnotesize{Total density of states (TDOS, blue) and projected density of states (PDOS) for the DTP backbone (orange) and TPA pendant groups (purple), along with the overlap population DOS (OPDOS, red), for the 
{\bf{(a)}} neutral, 
{\bf{(b)}} monooxidized $+1$, 
{\bf{(c)}} dication singlet $+2$, 
{\bf{(d)}} and dication triplet $+2$ 
states of the \activepoly\ pentamer.
Computed at B3LYP-D3BJ/def2-TZVP level in the gas phase. 
}}
\label{fig:DOS_combine_01}
\end{figure*}

By contrast, the open-shell singlet ($S=0$) exhibits degenerate HOMOs on the TPA side-chains ($E=-6.74$ to $-6.78~\mathrm{eV}$), whereas its LUMO ($E=-6.63~\mathrm{eV}$) is predominantly on the DTP backbone with partial delocalization onto the TPA pendants---resulting in a gap of $0.11~\mathrm{eV}$. These spin-dependent orbital distributions highlight a bifurcated, spin-resolved percolation network: in the triplet state, charge-transport pathways on TPA (HOMOs) and DTP (LUMOs) are spatially segregated, whereas in the singlet diradical the LUMO acquires mixed backbone/pendant character. 

All calculations employed the B3LYP-D3BJ functional with the def2-TZVP basis set in the gas phase. The DFT-computed HOMO ($-4.27~\mathrm{eV}$) and LUMO ($-2.19~\mathrm{eV}$) levels of the neutral pentamer in the gas phase are in good agreement with the previously presented 
experimental CV/UV-Vis values determined for polymer films on ITO ($-4.8~\mathrm{eV}$ and $-2.0~\mathrm{eV}$, respectively) as well
as with ER-EIS measurements ($-4.67~\mathrm{eV}$ and $-1.88~\mathrm{eV}$). 
The TD-DFT optical gap of $2.08~\mathrm{eV}$ likewise compares favorably with the experimental gap ($2.79$--$2.83~\mathrm{eV}$), differing by $0.7~\mathrm{eV}$. 
This close alignment ($\Delta \approx 0.3$--$0.7~\mathrm{eV}$) between theory and experiment supports the conclusion that the DTP--TPA system exhibits suitably aligned frontier orbitals for efficient charge injection and extraction in memristor devices despite systematic shifts inherent to gas-phase modeling.

{\bf{Figure~\ref{fig:DOS_combine_01}}} presents the total (TDOS) and 
fragment-projected (PDOS) densities of states for the DTP backbone and TPA pendants, 
along with the overlap population DOS (OPDOS), for the neutral, monooxidized ($+1$), and dication ($+2$) states. 
In the neutral state (cf.\ {\bf{Figure \ref{fig:DOS_combine_01}(a)}}), 
both the HOMO and LUMO are localized on DTP, consistent with the spin-density plot showing no unpaired electron density on TPA. 
Upon first oxidation (cf.\ {\bf{Figure \ref{fig:DOS_combine_01}(b)}}), the HOMO shifts to TPA while the LUMO remains on DTP, narrowing the gap to $\approx 0.10~\mathrm{eV}$; this aligns with the radical-cation spin-density plot showing the unpaired electron confined to DTP. 
In the singlet dication (cf.\ {\bf{Figure \ref{fig:DOS_combine_01}(c)}}), the HOMO energies of DTP and TPA become similar, with the LUMO gaining partial TPA character, while in the triplet dication (cf.\ {\bf{Figure \ref{fig:DOS_combine_01}(d)}}), 
HOMOs remain fully on TPA and LUMOs on DTP. 
Across all oxidation states, the OPDOS remains near zero, indicating minimal direct orbital overlap between backbone and pendant groups. 

The voltage dependence of the discrete Shannon entropy 
(cf.\ {\bf Figure~\ref{fig:entropy}}) can be directly related to the orbital energetics revealed by DFT 
(cf.\ {\bf Figure~\ref{fig:DOS_combine_01}}). Projected DOS analysis shows that in the neutral state both the HOMO and LUMO are localized on the DTP backbone, while oxidation progressively activates transport channels on the TPA pendants. In the dication state, the HOMO levels of DTP and TPA become nearly degenerate, yet remain spatially separated as indicated by the near-zero OPDOS. This near-degeneracy balances the backbone- and pendant-localized states, creating a bias regime in which stochastic switching between conduction pathways is maximized. The resulting nearly equiprobable output distribution corresponds to the entropy peak observed near $\pm 1~\mathrm{V}$. At higher or lower biases, the applied field shifts the relative alignment of the frontier orbitals, favoring one pathway and reducing entropy. Thus, the experimentally observed information maximum is a direct manifestation of the polymer’s electronic structure.

\section{Conclusion}

We have shown that pTPAC6DTP memristors operate as intrinsic entropy sources capable of generating voltage-gated probabilistic bits. Peaks in discrete Shannon entropy from randomly pulsed $I$--$V$ measurements coincide with the bias window where circuit-level transfer functions exhibit maximal fluctuations, confirming a shared physical origin in the nonlinear memristive response. Dielectric spectroscopy demonstrates that the pendant TPA units are dynamically active, providing temperature- and field-driven variability that modulates percolation pathways in real time. Energy-resolved electrochemical and quantum-chemical analyses further reveal that the energetics of TPA, DTP, and ITO align within the transport gap, producing a bifurcated percolation network that is tunable by molecular design and biasing conditions.  

These results establish a direct molecular-to-device connection between microscopic relaxation processes, energetic alignment, and circuit-level stochasticity. By harnessing, rather than suppressing, the intrinsic fluctuations of the TPA pendants, \activepoly\ devices embody the central principle of thermodynamic computing, computation driven by controlled thermal noise. This work therefore positions organic memristors as the first structurally tunable polymeric building blocks for scalable probabilistic hardware.

\section{Experimental}

\subsection{Materials and methods}

All reagents were obtained from commercial suppliers (TCI America, Fisher Scientific, and VWR) and used as received unless otherwise noted. Reaction solvents were dried with standard agents, distilled under argon, and employed immediately. NMR spectra ($^{1}$H and $^{13}$C) were collected on a JEOL ECX–300 instrument. Proton chemical shifts are reported in ppm relative to tetramethylsilane, with referencing to the residual protio signal of CDCl$_3$ ($\delta$ 7.26 ppm). Carbon chemical shifts are given relative to tetramethylsilane and referenced to the solvent peak of CDCl$_3$ ($\delta$ 77.16 ppm). Melting points were measured on an EZ-Melt automated apparatus. High-resolution mass spectra were obtained on a Waters Q-ToF Premier instrument.
				
\subsection{Synthesis}

\begin{figure}[h]
\begin{center}
\scalebox{0.45}[0.45]{\includegraphics{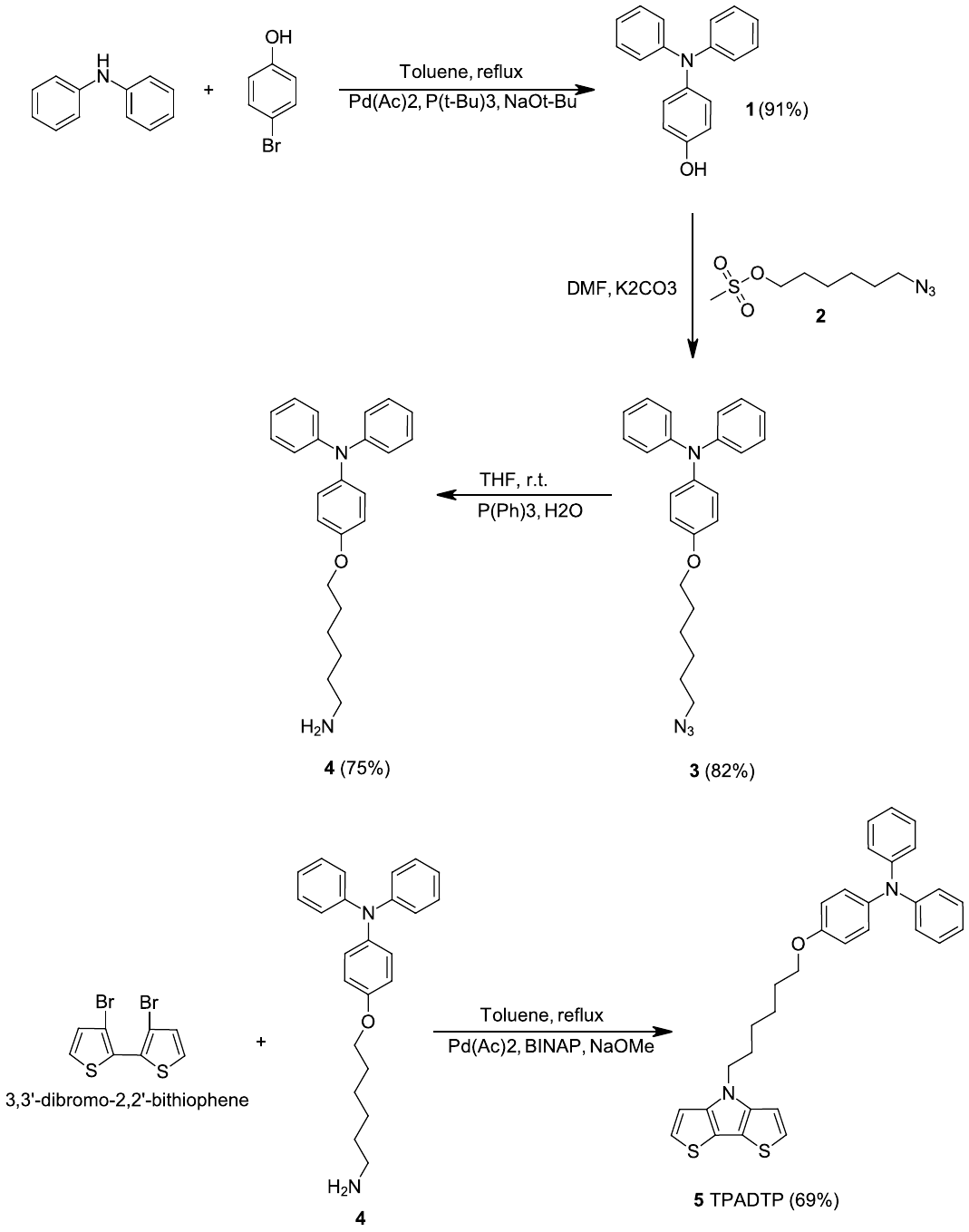}}
\caption{\footnotesize{Synthetic scheme for the preparation of the monomer
4-((6-(4H-dithieno[3,2-b:2',3'-d]pyrrol-4-yl)hexyl)oxy)-N,N-diphenylaniline (TPAC$_6$DTP), illustrating the key reaction steps and intermediates leading to the final product.}}
\label{fig:scheme}
\end{center}
\end{figure}

Refer to {\bf{Figure \ref{fig:scheme}}} for the synthetic procedure.

\subsubsection{4-(Diphenylamino)phenol (1)}  
A round-bottom flask was charged under nitrogen with sodium \emph{tert}-butoxide (1.7 g, 17.7 mmol), diphenylamine (1.5 g, 8.86 mmol), \emph{p}-bromophenol (1.65 g, 8.8 mmol), and palladium(II) acetate (10 mg, 0.04 mmol). A solution of tri-\emph{tert}-butylphosphine (40 mg, 0.2 mmol) in toluene (30 mL) was then added, and the mixture was stirred under nitrogen at 100~$^\circ$C for 5 h. After cooling to room temperature, the reaction was diluted with ethyl acetate (20 mL) and washed with water. The organic layer was dried over Na$_2$SO$_4$, filtered, and concentrated. Purification by column chromatography on silica (dichloromethane/hexane, 2:1) afforded a colorless solid (2.1 g, 91\%, m.p. 118–119~$^\circ$C). $^1$H NMR (DMSO-$d_6$): $\delta$ 6.75 (d, 2H, $J=8.9$ Hz), 6.92 (m, 8H), 7.22 (m, 4H, $J=8.9$ Hz).  

\subsubsection{6-Azidohexyl Methanesulfonate (2)}  
The compound was prepared following a previously reported method\cite{shon1999}.  

\subsubsection{4-((6-azidohexyl)oxy)-\emph{N,N}-diphenylaniline (3)}  
4-(Diphenylamino)phenol (2 g, 7.65 mmol) and 6-azidohexyl methanesulfonate (2.2 g, 9.95 mmol) were dissolved in dimethylformamide (40 mL), and potassium carbonate (1.37 g, 9.95 mmol) was added. The mixture was degassed with nitrogen and stirred at 80~$^\circ$C for 4 h. After cooling, the mixture was extracted with dichloromethane and washed three times with water. The combined organic extracts were evaporated, and the residue was purified by column chromatography (dichloromethane/hexane, 1:1) to give a clear oil (2.42 g, 82\%). $^1$H NMR (CDCl$_3$): $\delta$ 1.48 (m, 4H), 1.64 (m, 2H, $J=6.5,7.2$ Hz), 1.80 (m, 2H, $J=6.5,7.2$ Hz), 3.29 (t, 2H, $J=7.2$ Hz), 3.94 (t, 2H, $J=6.5$ Hz), 6.82 (d, 2H, $J=8.6$ Hz), 6.94 (m, 2H, $J=7.2,8.6$ Hz), 7.04 (m, 6H, $J=8.6$ Hz), 7.21 (m, 4H, $J=7.2$ Hz).  

\subsubsection{4-((6-aminohexyl)oxy)-\emph{N,N}-diphenylaniline (4)}  
Compound 3 (1.8 g, 4.65 mmol) was dissolved in tetrahydrofuran (15 mL), and triphenylphosphine (1.34 g, 5.12 mmol) was added. The solution was stirred at room temperature for 16 h, followed by addition of water (3 mL) and continued stirring for 3 h. The solvent was removed under vacuum, and the residue was dissolved in methanol and treated with concentrated HCl (35\%, 1 mL). After evaporation, the residue was suspended in water (80 mL). The insoluble material was separated by centrifugation, and the aqueous phase containing the hydrochloride salt was basified with sodium carbonate. The free amine was extracted with dichloromethane, and the organic phase was dried, filtered, and concentrated. The crude product was purified by silica gel chromatography: impurities were eluted with dichloromethane/methanol (3:1), and the target compound was collected using the same mixture containing 4\% ammonium hydroxide solution. The product was obtained as a pale yellow oil (1.25 g, 75\%). $^1$H NMR (CDCl$_3$): $\delta$ 1.47 (m, 6H), 1.79 (m, 2H, $J=6.5$ Hz), 2.73 (m, 2H, $J=6.9$ Hz), 3.93 (t, 2H, $J=6.5$ Hz), 6.81 (d, 2H, $J=8.9$ Hz), 6.93 (m, 2H, $J=7.2$ Hz), 7.04 (m, 6H, $J=8.9$ Hz), 7.20 (m, 4H, $J=7.2$ Hz).  

\subsubsection{4-((6-(4H-dithieno[3,2-b:2’,3’-d]pyrrol-4-yl)hexyl)oxy)-\emph{N,N}-diphenylaniline (5)}  
3,3'-Dibromo-2,2'-bithiophene (0.78 g, 2.4 mmol) and compound 4 (0.87 g, 2.4 mmol) were dissolved in dry toluene (35 mL) and degassed with nitrogen. This solution was transferred into a flask containing BINAP (80 mg), palladium(II) acetate (15 mg), and sodium methoxide (0.39 g, 7.2 mmol) under nitrogen. The reaction mixture was refluxed with stirring for 30 h, cooled, diluted with ethyl acetate (30 mL), and washed with water. The organic layer was collected, concentrated, and purified by column chromatography (dichloromethane/hexane, 1:1) to give a clear oil (0.86 g, 69\%). $^1$H NMR (CDCl$_3$): $\delta$ 1.37–1.53 (m, 4H), 1.74 (m, 2H, $J=6.5$ Hz), 1.91 (m, 2H, $J=6.9$ Hz), 3.88 (t, 2H, $J=6.5$ Hz), 4.22 (t, 2H, $J=6.9$ Hz), 6.79 (d, 2H, $J=8.6$ Hz), 6.94 (m, 2H, $J=7.2$ Hz), 7.00–7.07 (m, 8H, $J=5.9,7.2,8.6$ Hz), 7.12 (d, 2H, $J=5.9$ Hz), 7.21 (m, 4H, $J=7.2$ Hz). $^{13}$C NMR (CDCl$_3$): $\delta$ 25.9, 26.9, 29.3, 30.5, 47.4, 68.0, 111.0, 114.8, 115.4, 121.9, 122.9, 123.0, 127.5, 129.2, 140.7, 145.0, 148.3, 155.7. ESI$^+$ mass spectrum: calculated for C$_{32}$H$_{30}$N$_2$OS$_2$ [M]$^+$ 522.18, found 522.179.

\subsection{Electropolymerization and device fabrication}

Device substrates for electropolymerization were prepared from unpolished float glass slides (12.7 mm × 12.7 mm × 0.7 mm) coated on one side with SiO\textsubscript{2}-passivated indium tin oxide (ITO, sheet resistance 8–12~$\Omega$). A 4 mm strip of vinyl tape was applied along the center of the ITO to define the anode area, and the exposed ITO regions were etched by covering with a thin layer of zinc powder and subsequently treating with concentrated HCl (36.5–38\%). Following etching, the slides were rinsed twice with deionized (DI) water, once with the tape in place and again after tape removal. Substrates were then cleaned sequentially by sonication in acetone (10 min) and isopropanol (10 min), with intermediate wiping using cotton swabs, before drying under nitrogen. A final plasma treatment was carried out in a Harrick PDC-32G plasma cleaner (high setting, 5 min) to eliminate organic residues and improve surface wettability.  

Cyclic voltammetry (CV) and electropolymerization were performed on a 
BASi 100A Electrochemical Analyzer with a C3 Cell Stand. 
For film growth, 0.5~mL of a 0.015~M solution of 
4-((6-(4H-dithieno[3,2-b:2',3'-d]pyrrol-4-yl)hexyl)oxy)-\textit{N,N}-diphenylaniline 
in acetonitrile (ACN) containing 0.1~M TBAPF\textsubscript{6} 
was combined with 9~mL of a 0.1~M TBAPF\textsubscript{6} ACN stock solution.
The resulting solution was transferred to a nitrogen-purged electrochemical cell equipped 
with an Ag/AgCl reference electrode, a platinum wire counter electrode, and the patterned 
ITO slide as the working electrode. After 5~min of stirring and nitrogen purging, 
electropolymerization was initiated by cycling the potential between –0.5~V and +0.9~V 
(vs.\ Ag/AgCl) at 100~mV/s for 25 cycles. Following the cycling, the films were held at 
+0.9~V (vs.\ Ag/AgCl) for 30~s in the monomer-containing solution to complete oxidation. 
The electrolyte was then replaced with monomer-free supporting electrolyte, and the films 
were again held at +0.9~V for 30~s to fully oxidize and stabilize the polymer prior to 
device assembly.
  
Residual salts and unreacted monomer were removed by rinsing the polymer-coated slides first with neat ACN and subsequently with DI water. The resulting electropolymerized films were bluish in appearance and had thicknesses between 100 and 300 nm.  

Aluminum top contacts were deposited using a Denton Vacuum DV-502A thermal evaporator. 
Aluminum pellets (Kurt J. Lesker, 1/4-inch diameter, 1/2-inch length) were evaporated from a tungsten basket coil under high vacuum (2 × 10\textsuperscript{–6} torr). 
Substrates were placed in a custom shadow mask designed to define two aluminum contacts orthogonal to the underlying ITO strip, producing two devices per slide with active areas of 4 mm\textsuperscript{2} each. 
The deposition was monitored with a Sigma Instruments SQM-160 rate and thickness controller, and aluminum layers of approximately 200 nm were obtained.

\subsection{Electrical characterization and p-bit generation}		

For the randomized $I$--$V$ response, the electrical measurements were conducted using an HP~4156A Semiconductor Parameter Analyzer coupled with an HP~16058A Test Fixture via triaxial leads, all controlled by customized Python code. The device was subjected to random voltages uniformly selected from the range $-3$~V to $+3$~V. Each voltage was applied as a 180~ms square-wave pulse, during which the resulting current was recorded, followed by a 100~ms grounding interval through both electrodes before the next voltage was applied. Unless otherwise specified, the ITO electrode was maintained at zero potential, while potentials were applied to the aluminum electrode.  

For analysis, the ensemble of measured currents at each voltage bin was converted into a normalized probability distribution $P(I\mid V)$ using 100 logarithmically spaced current bins spanning $10^{-15}$--$10^{-2}$~A. From these distributions, the discrete Shannon entropy was calculated 
using {\bf{Eq.\ \ref{eq:disc_entropy}}}.
The effective number of equiprobable states was then obtained as $N_{\mathrm{eff}}(V)=2^{H_{\mathrm{disc}}(V)}$.
Because $H_{\mathrm{disc}}$ is computed from binned continuous data, its absolute value depends on the number of bins, the current range considered,  the choice of logarithmic spacing, and, consequently, values can exceed 1~bit.

The circuit presented in {\bf{Figure~\ref{fig:circuit}}} was implemented on an 
HP~4156A Semiconductor Parameter Analyzer with an HP~16442A test fixture, 
operated under Python control, and configured to generate probabilistic bits 
(p-bits) using the \activepoly\ memristor in a purely voltage-driven mode. 
The design consists of a voltage divider followed by a comparator-based 
thresholding stage, with the supply voltage ($V_{DD}$) varied to examine its 
influence on the system response parameter $k$. The polymer memristor 
($R_{\text{mem}}$) occupies the upper branch of the divider and a 
100~k$\Omega$ resistor ($R_{s}$) the lower branch. Adjusting $V_{DD}$ 
modulates the divider midpoint ($V_{\text{mvd}}$), thereby tuning the 
probability distribution of the p-bit output. A 741 operational amplifier 
buffers $V_{\text{mvd}}$, isolating the divider from downstream circuitry 
and preventing loading. The buffered signal is compared against a reference 
voltage ($V_{\text{ref}}$) by an LM393 comparator, which generates a TTL 
output depending on whether $V_{\text{mvd}}$ lies above or below $V_{\text{ref}}$. 
Comparator stability is maintained by a 10~k$\Omega$ pull-up resistor ($R_{l}$) 
and a 10~pF capacitor ($C_{l}$), ensuring clean switching. To account for slow 
drift in the memristor response, $V_{\text{ref}}$ was periodically updated 
every 10 cycles. Unless otherwise specified, the memristor was biased for hole 
transport, with the aluminum electrode held at lower potential relative to the 
ITO electrode.

\subsection{Thermal characterization}
For dielectric analysis, a TA Instruments DEA 2970 Dielectric Analyzer operating in the 
parallel-plate configuration was used to investigate the thermal relaxation behavior of 
\activepoly. Following the procedure established for electropolymerizing the polymer onto 
ITO substrates, films were instead deposited directly onto the gold guard-ring electrodes 
of the DEA cell. 
After electropolymerization, the films were conditioned by applying a 
reduced bias voltage for 2~min to ensure that they were in a non-conductive state, producing 
a uniform coating of the reduced polymer on the electrodes. 
To minimize residual electrolyte 
and remove absorbed salt, the electrodes were immersed in a monomer-free, salt-free 
acetonitrile solution and held at –0.8~V (vs.\ Ag/AgCl)  for 3~h, then briefly transferred to a hot plate at 
100~$^{\circ}$C for 10~min to dry. 
The sensors were subsequently placed under a continuous nitrogen 
purge in the DEA for 24~h prior to measurement. This desalting and conditioning sequence 
was essential for maximizing the insulating character of the films, thereby preventing ionic 
conduction from overwhelming the dielectric response associated with molecular dipole 
mobility. Measurements were carried out under nitrogen purge with temperature ramped from 
–150~$^{\circ}$C to +120~$^{\circ}$C at 1~$^{\circ}$C/min, using a frequency sweep from 
0.01~Hz to 100~kHz.

\subsection{Theoretical calculations}

Ground-state geometries of neutral oligomers comprising 2 to 10 repeat units in an antiperiplanar arrangement were constructed and optimized in the gas phase using the GFN2-xTB method (xtb~6.6.1).\cite{RN187,RN166} 
Single-point energy calculations were then carried out on the 2-, 4-, 6-, and 8-unit oligomers at the B3LYP-D3BJ/def2-TZVP level to evaluate their frontier orbital energies. 
For the pentamer, geometries of the neutral, radical cation, triplet diradical dication, and singlet dication states were optimized in the gas phase using the TPSS meta-GGA functional with D3BJ dispersion correction and the def2-SVP basis set, as implemented in ORCA~6.0.1. 
Subsequently, single-point energy and molecular orbital calculations were carried out on these geometries at the B3LYP-D3BJ/def2-TZVP level.\cite{RN269,RN33,RN75} 
Unrestricted (UHF) calculations were employed for the doublet and triplet spin states, while restricted calculations were used for singlets. 
Adiabatic ionization potentials were obtained as total energy differences between the optimized charged and neutral species.\cite{SAFI2022139349} 
Vertical excitation energies were computed via time-dependent DFT (TD-DFT) at the B3LYP-D3BJ/def2-TZVP level in the gas phase.\cite{RN31}

\begin{acknowledgement}
The authors thank the Gregg-Graniteville Foundation,the National Science Foundation (OIA-1632881 \& OIA-1655740), the Czech Science Foundation (24-10384S), and Ministry of Education, Youth and Sports of the Czech Republic Program INTER-EXCELLENCE (LUAUS24032) for financial support.
\end{acknowledgement}

\bibliography{Memory_020425,11092025_DFT_ref}

\end{document}